\documentclass[prx,aps,twocolumn,reprint,amsmath,floatfix,amssymb,showpacs,superscriptaddress]{revtex4-1}

\usepackage{xfrac}
\usepackage{amsmath}
\usepackage{booktabs}
\usepackage{array}

\def\rot#1{\rotatebox{90}{#1}}

\usepackage{microtype}
\usepackage{wasysym}
\usepackage[varg]{txfonts}
\usepackage[percent]{overpic}
\usepackage{mathrsfs}
\usepackage{braket}
\usepackage{titlesec}
\usepackage{bm}
\usepackage{comment}
\usepackage{verbatim}

\usepackage{array}
\newcommand{\PreserveBackslash}[1]{\let\temp=\\#1\let\\=\temp}
\newcolumntype{C}[1]{>{\PreserveBackslash\centering}p{#1}}
\newcolumntype{R}[1]{>{\PreserveBackslash\raggedleft}p{#1}}
\newcolumntype{L}[1]{>{\PreserveBackslash\raggedright}p{#1}}

\usepackage{color}
\usepackage[colorlinks,bookmarks=false,citecolor=blue,linkcolor=blue,urlcolor=blue,pdfstartview=FitH]{hyperref}
\def\cred{\color{red}}
\definecolor{darkred}{rgb}{0.7,0.0,0.0}

\definecolor{darkblue}{rgb}{0,0.02,0.45}

\definecolor{darkgreen}{rgb}{0.02,0.45,0.0}

\definecolor{violet}{rgb}{0.8,0.2,0.6}

\usepackage{amsmath}
\usepackage{amssymb}
\usepackage{graphicx}
\usepackage{subfigure}
\usepackage{pict2e}
\usepackage{multirow}
\newcommand{\be}{\begin{equation}}
\newcommand{\ee}{\end{equation}}
\newcommand{\bea}{\begin{eqnarray}}
\newcommand{\eea}{\end{eqnarray}}
\newcommand{\sbe}{\small\begin{equation}}
\newcommand{\see}{\end{equation}\normalsize}
\newcommand{\sbea}{\small\begin{eqnarray}}
\newcommand{\seea}{\end{eqnarray}\normalsize}
\graphicspath{{Figures/}}

\def\bs{\boldsymbol}
\def\vec{\mathbf}
\def\mc{\mathcal}
\def\nn{\nonumber}
\begin{document}

\def\Ha{${\bf H}\!\parallel\!{\bf a}$}
\def\Hb{${\bf H}\!\parallel\!{\bf b}$}
\def\Hc{${\bf H}\!\parallel\!{\bf c}$}

\title{Unconventional magnetic field response of the hyperhoneycomb Kitaev magnet $\beta-\textbf{Li}_2\textbf{IrO}_3$}

\author{Mengqun Li}
\affiliation{School of Physics and Astronomy, University of Minnesota, Minneapolis, MN 55455, USA}

\author{Ioannis Rousochatzakis}
\affiliation{Department of Physics, Loughborough University, Loughborough LE11 3TU, United Kingdom}

\author{Natalia B. Perkins}
\affiliation{School of Physics and Astronomy, University of Minnesota, Minneapolis, MN 55455, USA}
\date{\today}

\begin{abstract}
We present a unified description of the response of the hyperhoneycomb Kitaev magnet $\beta$-$\text{Li}_2\text{IrO}_3$ to applied magnetic fields along the orthorhombic directions ${\bf a}$, ${\bf b}$ and ${\bf c}$. This description is based on the minimal nearest-neighbor $J$-$K$-$\Gamma$ model and builds on the idea that the incommensurate counter-rotating order observed experimentally at zero field can be treated as a long-distance twisting of a nearby commensurate order with six spin sublattices. 
The results reveal that the behavior of the system for {\Ha}, {\Hb}  and {\Hc} share a number of qualitative  features, including: i) a strong intertwining of the modulated, counter-rotating order with a set of uniform orders; ii) the disappearance of the modulated order at a critical field $H^\ast$, whose value is strongly anisotropic with $H_{\bf b}^\ast\!<\!H_{\bf c}^\ast\!\ll\!H_{\bf a}^\ast$; iii) the presence of a robust zigzag phase above $H^\ast$; and iv) the fulfillment of the Bragg peak intensity sum rule. 
It is noteworthy that the disappearance of the modulated order for {\Hc} proceeds via a `metamagnetic' first-order transition which does not restore all broken symmetries. This implies the existence of a second finite-$T$ phase transition at higher magnetic fields. 
We also demonstrate that quantum fluctuations give rise to a significant reduction of the local moments for all directions of the field. The results for the total magnetization for {\Hb} are consistent with available data and confirm a previous assertion that the system is very close to the highly-frustrated $K$-$\Gamma$ line in parameter space. Our predictions for the magnetic response for fields along ${\bf a}$ and ${\bf c}$ await experimental verification. 
\end{abstract}

\maketitle

\vspace*{-1cm}
\section{Introduction}
\vspace*{-0.3cm}
In recent years there has been a  growing interest in the magnetic properties of 4d and 5d transition metal compounds with tri-coordinated lattices and bond-directional exchange anisotropies, broadly known as Kitaev materials~\cite{Jackeli2009,Jackeli2010,BookCao,Rau2016,Trebst2017,Knolle2017,Winter2017,Takagi2019,Motome2019}. Among these, the most extensively studied are the iridates A$_2$IrO$_3$ (A = Li, Na)~\cite{Singh2010,Liu2011, Singh2012,Ye2012,Biffin2014a,Biffin2014b,Takayama2015,Chun2015, Williams2016,Modic2014,Breznay2017,Ruiz2017,Veiga2017,Takayama2019,Majumder2019,Majumder2019b} and H$_3$LiIr$_2$O$_6$~\cite{Kitagawa2018,Pei2019}, and the ruthenate $\alpha$-RuCl$_3$~\cite{Plumb2014, Sears2015, Majumder2015, Johnson2015}. The main interest in these materials has been triggered by the realization~\cite{Jackeli2009,Jackeli2010} that the dominant exchange interaction between the effective spin-orbit-entangled $j_{\rm eff}\!=\!1/2$ moments is the so-called Kitaev anisotropy which is known to stabilize a variety of quantum spin liquid phases~\cite{Kitaev2006,Mandal2009,Kimchi2014,Hermanns2016,IoannisClassKitaev}.

Besides the dominant Kitaev anisotropy, the above materials feature additional weaker interactions which generally give rise to a wealth of nontrivial phases competing with the quantum spin liquids~\cite{Jackeli2009,Jackeli2010,BookCao,Rau2016,Trebst2017,Knolle2017,Winter2017,Takagi2019}.
A central goal in the field is therefore to map out the various instabilities and identify the distinctive experimental signatures of the most relevant interactions. One of the most promising ways to achieve this goal experimentally is to subject the Kitaev materials in external magnetic fields along different directions. 
Apart from controlling the interplay of various zero-field competing phases, such external fields can also stabilize new collective spin states. There are, for example, many reports for possible magnetic field-induced quantum spin liquids~\cite{Baek2017,Zheng2017,Wolter2017,Kasahara2018}, and a variety of complex multi-sublattice, single- and multi-${\bf Q}$ phases~\cite{Janssen2016,Chern2017,Janssen2017,LiChern2019,Janssen2019}.

Remarkably, all experimental data reported so far  for Kitaev materials show that their response to the magnetic field depends very strongly on its direction. This is true for the layered compounds Na$_2$IrO$_3$~\cite{Singh2010}, $\alpha$-Li$_2$IrO$_3$~\cite{Freund2016}, and $\alpha$-RuCl$_3$~\cite{Sears2015,Johnson2015,Majumder2015,Kelley2018}, as well as for the three-dimensional (3D) iridates $\beta$-Li$_2$IrO$_3$~\cite{Ruiz2017,Majumder2019} and $\gamma$-Li$_2$IrO$_3$~\cite{Modic2014,Modic2018}. 
Here we revisit the case of the hyper-honeycomb $\beta$-Li$_2$IrO$_3$ and show that its strongly anisotropic response signifies a large separation of energy scales between the relevant microscopic interactions, and can thus be used to extract information about the relative strength of these interactions in a direct way.

The main features of $\beta$-Li$_2$IrO$_3$ that are known so far are as follows~\cite{Biffin2014a,Takayama2015,Veiga2017,Ruiz2017,Majumder2019}. 
At zero field, the system orders magnetically below $T_N\!=\!38$~K, with the spins forming a non-coplanar, incommensurate (IC) modulation, with propagation wavevector ${\bf Q}\!=\!(0.57,0,0)$ in the orthorhombic frame, and two counter-rotating sets of moments~\cite{Biffin2014a}, similar to those in $\gamma$-Li$_2$IrO$_3$~\cite{Biffin2014b} and $\alpha$-Li$_2$IrO$_3$~\cite{Williams2016}.
A magnetic field along ${\bf b}$ destroys the IC order at a characteristic field $H_{\bf b}^\ast\sim2.8$~T, beyond which the spins show a uniform ${\bf Q}\!=\!0$ coplanar phase, comprising a ferromagnetic (FM) component along the field and a robust zigzag component along ${\bf a}$~\cite{Ruiz2017}. These components are also present below $H_{\bf b}^\ast$, but are too small to be detected at zero field~\cite{Ducatman2018,Rousochatzakis2018}. 
For {\Ha} and {\Hc}, the system shows a much weaker response, with the IC order remaining robust and the magnetization being linear up to the maximum fields measured (see supplemental material in \cite{Ruiz2017} and also \cite{Majumder2019}).

On the theory side, it has been established that the magnetism of $\beta$-Li$_2$IrO$_3$ can be accurately described by the nearest neighbor (NN) $J$-$K$-$\Gamma$ model~\cite{Lee2015,Lee2016,Ducatman2018,Rousochatzakis2018}, where $K$ denotes the Kitaev coupling, $J$ the Heisenberg coupling and $\Gamma$ the so-called symmetric exchange anisotropy which is present in many Kitaev materials~\cite{Katukuri2014,Rau2014,Lee2015,Lee2016,KimKimKee2016,IoannisGamma}.
In particular, $\beta$-Li$_2$IrO$_3$ is believed to be in the regime of large negative $K$, large negative $\Gamma$ (with $|\Gamma|\!<\!|K|)$ and small positive $J$ (with $J\!\ll|\Gamma|$), see detailed discussion in \cite{Ducatman2018,Rousochatzakis2018}. 
Remarkably,  in this parameter regime, the critical field $H_{\bf b}^\ast$ depends only on $J$, specifically~\cite{Rousochatzakis2018} $\mu_B H_{\bf b}^\ast\!\sim\!0.46 J \left(4S/g_{bb}\right)$, where $S\!=\!1/2$ denotes the classical spin length of the $j_{\text{eff}}\!=\!1/2$ degree of freedom, $\mu_B$ is the Bohr magneton and $g_{bb}$ is the diagonal element of the electronic ${\bf g}$-tensor along ${\bf b}$. The small value of the experimentally measured  $H_{\bf b}^\ast$ is therefore a signature of the smallness of $J$ ($J\!\sim\!4$~K). 

It has also been shown~\cite{Ducatman2018,Rousochatzakis2018} that the IC order of $\beta$-Li$_2$IrO$_3$ can be treated as a long-distance twisting of a nearby commensurate period-3 state with ${\bf Q}\!=\!\frac{2}{3}\hat{\bf a}$ (in units $\frac{2\pi}{a}$). This state is amenable to a semi-analytical treatment of the problem, with results that are consistent with almost all experimental findings so far, both in zero field and at finite fields along ${\bf b}$~\cite{Ducatman2018,Rousochatzakis2018}. This analysis explains, for example, the presence of a uniform zigzag component along ${\bf a}$ on top of the modulated order, and the intensity sum rule of the corresponding Bragg peaks~\cite{Ruiz2017}. 

Here we show that this semi-analytical description can be naturally extended to the cases where ${\bf H}$ is along ${\bf a}$ and ${\bf c}$. The results, which are cross-checked with classical Monte Carlo simulations, show that the response along ${\bf a}$ and ${\bf c}$ directions shares many qualitative features with that along ${\bf b}$. Specifically, we find that the period-3 order disappears at a critical field $H^\ast$, whose value depends strongly on the field direction. Importantly, none of the critical fields depends on the Kitaev interaction $K$, and moreover $H_{\bf a}^\ast$ are mainly controlled by $\Gamma$. 
A realistic set of coupling parameters, 
\sbe\label{eq:JKGammaPars}
J\!=\!0.4~\text{meV}, ~~
K\!=\!-18~\text{meV} ~~ \text{and}~~
\Gamma\!=\!-10~\text{meV}\,
\see
delivers $H_{\bf b}^\ast\!\sim\!2.88$~T and $T_N\!\sim\!35.5$~K (in good agreement with  corresponding experimental values of 2.8~T and 38~K~\cite{Biffin2014a,Ruiz2017}), and also gives $H_{\bf a}^\ast\!\sim\!102$~T and $H_{\bf c}^\ast\!\sim\!13$~T. This means that at least the transition at $H_{\bf c}^\ast$ should be accessible experimentally, and the measured value of $H_{\bf c}^\ast$ can provide the value of $\Gamma$. 

The same semi-analytical approach provides a number of additional qualitative findings: 
(i) The period-3 order is always intertwined with a set of uniform orders, some of which give rise to a finite torque that can be measured experimentally. 
(ii) Among these uniform orders, there is always a zigzag component which remains robust above $H^\ast$ and coexists with the FM order along the field. Classically, the zigzag component disappears (but only for $g_{ab}\!=\!0$, see below) at $H^{\ast\ast}\!\to\!\infty$ for fields along ${\bf a}$ and ${\bf b}$, but for {\Hc} the corresponding field $H_{\bf c}^{\ast\ast}$ is finite. In particular, $H_{\bf c}^{\ast\ast}$ is governed mostly by $\Gamma$, with $H_{\bf c}^{\ast\ast}\!\sim\!45$~T for the parameters of Eq.~(\ref{eq:JKGammaPars}). 
(iii) The intensity sum rule between the Bragg peaks at ${\bf Q}\!=\!\frac{2}{3}\hat{\bf a}$ and ${\bf Q}\!=\!0$, which has been observed experimentally for {\Hb}~\cite{Ruiz2017}, is actually fulfilled for all field directions. As it turns out, this rule is an experimental fingerprint of the spin length constraints. 
(iv) While the transitions at $H_{\bf a}^\ast$ and $H_{\bf b}^\ast$ are continuous, the transition at $H_{\bf c}^\ast$ is of first order. Moreover, this transition does not restore all broken symmetries, which leads to the prediction of a second thermal phase transition at high enough fields along ${\bf c}$. This transition will be demonstrated explicitly by classical Monte Carlo simulations. 

One shortcoming of our semi-analytical classical approach is that it overestimates the magnetization at $H^\ast_{\bf b}$ by approximately a factor of two compared to the experimental value. This has led to the assertion~\cite{Rousochatzakis2018} that the spin lengths are strongly renormalized by quantum fluctuations due to the close proximity to the special $K$-$\Gamma$ line in parameter space, where the system is highly-frustrated~\cite{Ducatman2018}. 
This assertion is now demonstrated explicitly by a semiclassical $1/S$ expansion. The results confirm that the magnetization correction can be as large as 50\%, and a direct comparison with published experimental data shows good agreement for all three field directions. 

\begin{figure}[!t]
{\includegraphics[width=0.9\linewidth]{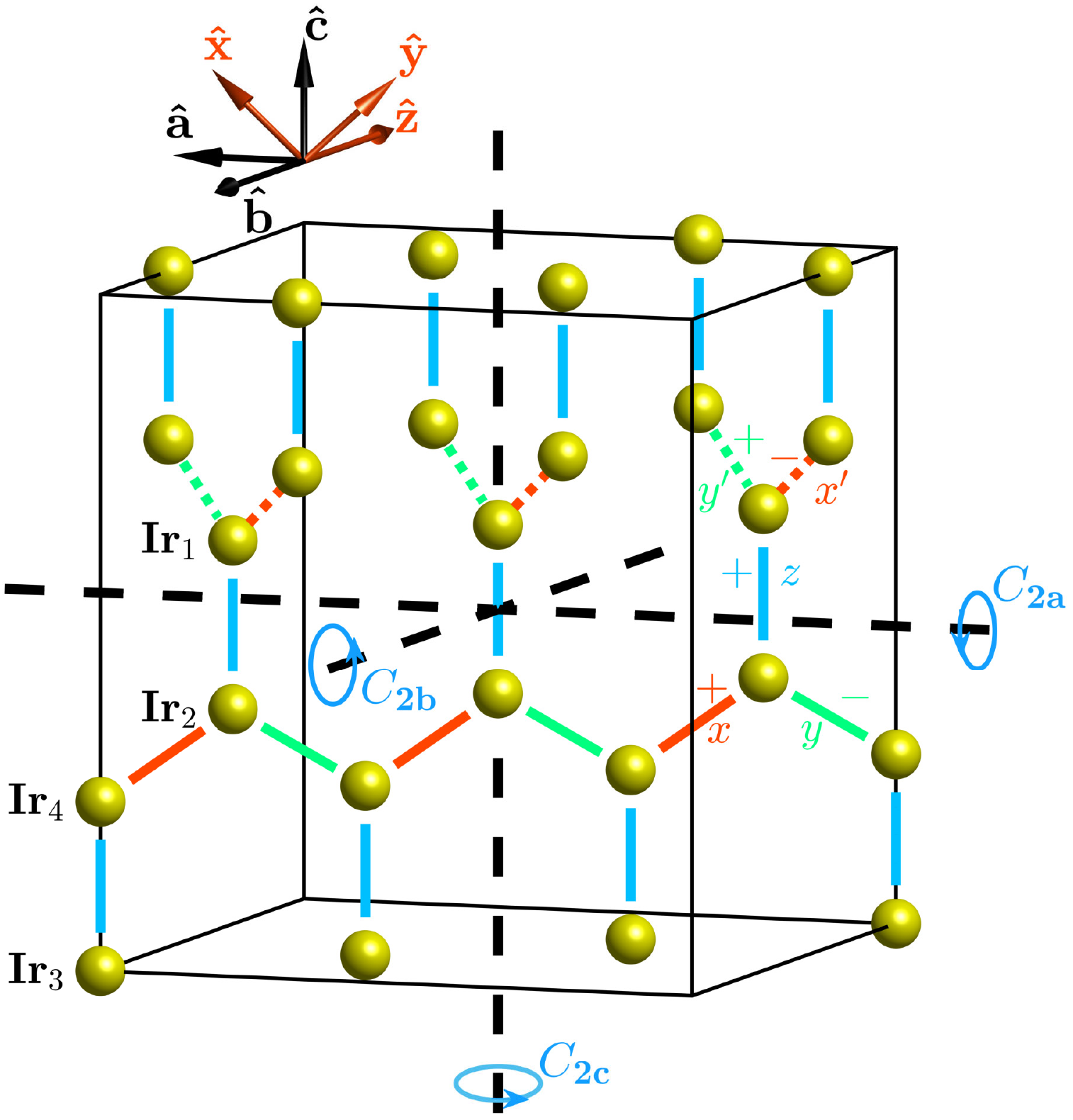}}
\caption{Sketch of a hyperhoneycomb lattice.  The five NN bonds of the $J$-$K$-$\Gamma$ model are shown by  solid (dashed) red lines for $t\in\{x,x'\}$,  solid (dashed) green lines for $t\in\{y,y'\}$, and solid blue lines for $t\in\{z\}$.}\label{fig:lattice} 
\end{figure}

The rest of the paper is organized as follows. In Sec.~\ref{sec:LatticeSymmetry}, we recall the structural and symmetry aspects of $\beta$-Li$_2$IrO$_3$ that are most relevant for this study. In Sec.~\ref{sec:Model}, we review the minimal $J$-$K$-$\Gamma$ model~\cite{Lee2015,Lee2016} and the symmetries of the corresponding spin Hamiltonian. In Sec.~\ref{sec:ansatze}, we present the unified semi-analytical description for all three orthorhombic directions. This includes the basic spin-sublattice structure of the various configurations (Sec.~\ref{sec:SpinSublattices}), their parametrization in terms of Cartesian components and the associated symmetry-resolved static structure factors (Secs.~\ref{sec:parametrization1}-\ref{sec:parametrization2}), the dependence of the critical fields $H^\ast$ on the model parameters (Sec.~\ref{sec:Hstar}), and a discussion of the symmetries that are broken in each regime (Sec.~\ref{sec:Symmetries}).
The role of quantum fluctuations is then addressed in Sec.~\ref{sec:QFs}, along with the direct comparison of the predicted magnetizations with available experimental data. 
In Sec.~\ref{sec:Torque} we discuss how the various transitions can be detected experimentally via  measurements of the magnetic torque, for which we provide predictions with and without harmonic spin-wave corrections. 
In Sec.~\ref{sec:FiniteT}, we cross-check our  ans\"atze with classical Monte Carlo simulations and compute the $H$-$T$ phase diagram for all three orthorhombic directions. Here we also highlight the qualitative difference between the zigzag orders for {\Ha} and {\Hb} versus the spontaneous high-field zigzag order for {\Hc}. 
A summary and a general discussion is given in Sec.~\ref{sec:Discussion}. 
Auxiliary information and technical details are provided in Appendices~\ref{app:SSF}-\ref{app:MC}.

\vspace*{-0.3cm}
\section{Lattice structure, symmetries \& conventions}\label{sec:LatticeSymmetry}
\vspace*{-0.3cm}
$\beta$-Li$_2$IrO$_3$ crystallizes in  a hyperhoneycomb structure (shown in Fig.~\ref{fig:lattice}) and has the F$ddd$ space group. Its conventional orthorhombic unit cell is set by the crystallographic axes $\{\hat{{\bf a}}, \hat{{\bf b}},\hat{{\bf c}}\}$, which are related to the Cartesian axes $\{\hat{{\bf x}}, \hat{{\bf y}}, \hat{{\bf z}}\}$ appearing in the spin Hamiltonian below [Eqs.~(\ref{eq:Hamiltonian0})-(\ref{eq:Hamiltonian})] by
\sbe\label{eq:xyzframe}
\hat{{\bf x}}=(\hat{{\bf a}}+\hat{{\bf c}})/\sqrt{2}\,,~~~
\hat{{\bf y}}=(\hat{{\bf c}}-\hat{{\bf a}})/\sqrt{2}\,,~~~
\hat{{\bf z}}=-\hat{{\bf b}}\,.
\see 
We note here that we stick to the $xyz$-frame convention of Refs.~[\onlinecite{Lee2015,Lee2016,Ducatman2018,Rousochatzakis2018}], which is different from the one used in Ref.~[\onlinecite{Biffin2014a}]. 
The two $xyz$-frames are related to each other by a two-fold rotation around the ${\bf x}$-axis. This is important as the choice of the frame affects the overall sign structure of the $\Gamma$ interactions. 

The orthorhombic unit cell contains four primitive unit cells, of four $\text{Ir}^{4+}$ ions each (labeled by Ir$_1$-Ir$_4$ in Fig.~\ref{fig:lattice}). 
The $\text{Ir}^{4+}$ ions form a hyperhoneycomb structure, which can be viewed as a stacking of two types of zigzag chains, which we will denote by $xy$- and $x'y'$-chains. The $xy$-chains run along the direction ${\bf a}\!+\!{\bf b}$ and are shown in Fig.~\ref{fig:lattice} by the alternating red and green solid bonds, denoted by $x$ and $y$ respectively. The $x'y'$-chains run along ${\bf a}\!-\!{\bf b}$ and are shown in Fig.~\ref{fig:lattice} by the alternating red and green dashed bonds, denoted by $x'$ and $y'$ respectively. The two types of chains are interconnected with vertical NN Ir-Ir bonds denoted in Fig.~\ref{fig:lattice} by $z$ (blue solid lines). In total, there are five types of NN Ir-Ir bonds, $x$, $y$, $x'$, $y'$ and $z$. 

Apart from translations, the crystal structure is invariant under the following point group operations~\cite{Ruiz2017}: (i) Inversion $\mc{I}$ through the center of every $x$- or $y$- or $x'$- or $y'$-type of bond, such as the center of the  Ir$_2$-Ir$_4$ bond of  Fig.~\ref{fig:lattice}. 
(ii) Three $\pi$-rotations in combined spin-orbit space, $C_{2\bf a}$, $C_{2\bf b}$, and $C_{2\bf c}$, around the axes ${\bf a}$, ${\bf b}$ and ${\bf c}$, respectively, passing through the middle of the $z$ bonds, as shown in Fig.~\ref{fig:lattice}. In particular, $C_{2\bf a}$ maps $x$-bonds to $y'$-bonds and $y$-bonds to $x'$-bonds in real space, and $[S_x,S_y,S_z]\rightarrow[-S_y,-S_x,-S_z]$ in spin space.  
Similarly, $C_{2\bf b}$ maps $x$-bonds to $x'$-bonds and $y$-bonds to $y'$-bonds in real space, and $[S_x,S_y,S_z]\rightarrow[-S_x,-S_y,S_z]$ in spin space. 
Finally, $C_{2\bf c}$ maps $x$-bonds to $y$-bonds and $x'$-bonds to $y'$-bonds in real space, and $[S_x,S_y,S_z]\rightarrow[S_y,S_x,-S_z]$ in spin space. 
(iii) Three glide planes which arise by reflections across the $\bf ab$-, $\bf bc$- and $\bf ac$-planes passing through an inversion center, followed by nonprimitive translations by $(\frac{1}{4}\frac{1}{4}0)$,  $(0\frac{1}{4}\frac{1}{4})$ and $(\frac{1}{4}0\frac{1}{4})$, in orthorhombic units, respectively.

At this point it is also worth introducing some terminology that we will need later  in the analysis of the static structure factors. Following Ref.~[\onlinecite{Biffin2014a}], we define four-component symmetry basis vectors, 
\sbe\label{eq:symmetrybasis}
A\!=\!\begin{pmatrix} 1\\-1\\-1\\1\end{pmatrix},~~~ 
C\!=\!\begin{pmatrix} 1\\1\\-1\\-1\end{pmatrix},~~~  
F\!=\!\begin{pmatrix}1\\1\\1\\1\end{pmatrix},~~~  
G\!=\!\begin{pmatrix}1\\-1\\1\\-1\end{pmatrix}\,.  
\see
These vectors represent, respectively, the relative amplitudes of the four sites of the primitive unit cell in the N\'eel (A), stripy (C), ferromagnetic (F) and zigzag (G) order.  
Note that, for consistency, our 4-site labeling Ir$_1$-Ir$_4$ of Fig.~\ref{fig:lattice} follows the convention of Fig.~7 of Ref.~[\onlinecite{Biffin2014a}].

For the various components of the static structure factor, we follow the convention of Ref.~[\onlinecite{Ducatman2018}] and denote the modulated components with ${\bf Q}\!=\!\frac{2}{3}\hat{{\bf a}}$ by the letter $M$ and the uniform components with ${\bf Q}\!=\!0$ by $M'$. Therefore, $M_a(A)$ denotes the modulated N\'eel ($A$) component along ${\bf a}$, $M'_b(F)$ denotes the uniform ferromagnetic ($F$) component along ${\bf b}$, and so on. The definitions of these components in terms of the Fourier transform of the spin configuration are given in Appendix~\ref{app:SSFconventions}.

\vspace*{-0.3cm}
\section{The minimal $J$-$K$-$\Gamma$ model} \label{sec:Model}
\begin{figure}
\includegraphics[width=0.7\columnwidth]{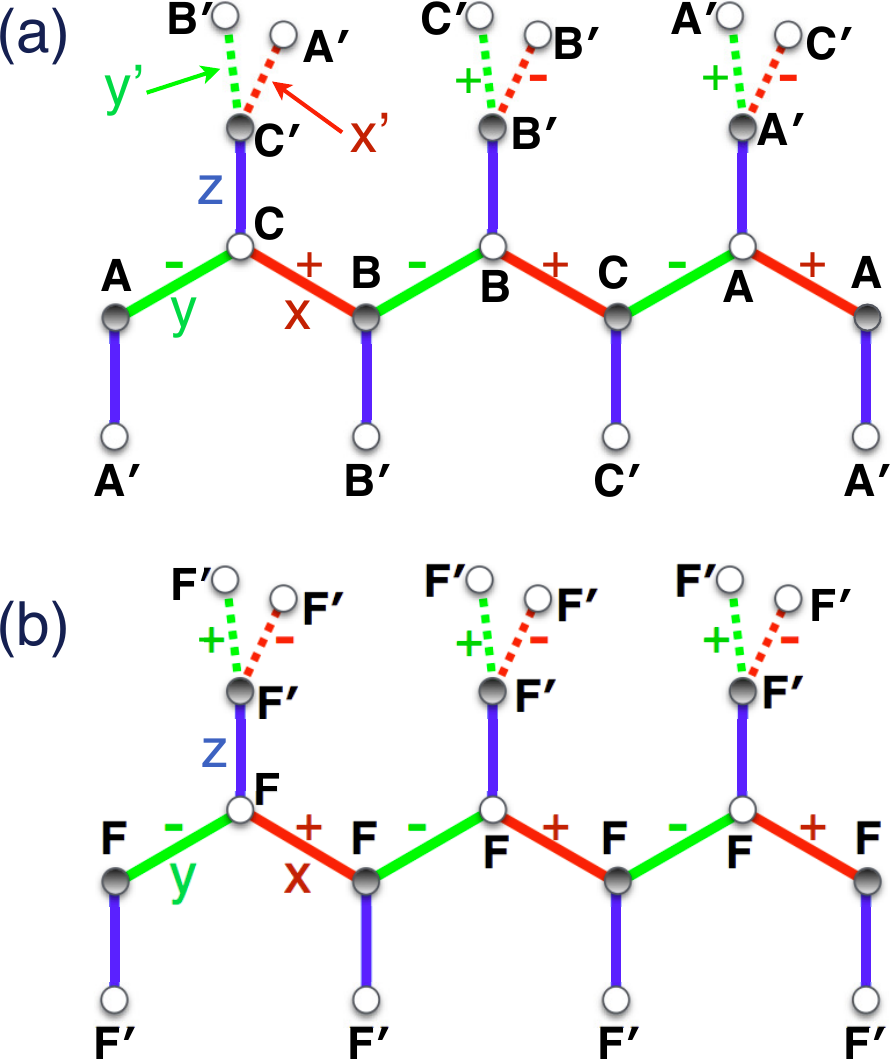}
\caption{General structure of the two field-induced phases of $\beta$-Li$_2$IrO$_3$ for ${\bf H}$ along ${\bf a}$, ${\bf b}$ or ${\bf c}$. (a) The six-sublattice low-field phase ($0\!\le\!H\!<H^\ast$). (b) The two-sublattice high-field phase ($H^\ast\!\le\!H\!<H^{\ast\ast}$), where $H_{{\bf a},{\bf b}}^{\ast\ast}\!=\!\infty$ and $H_{\bf c}^{\ast\ast}$ is finite. The Cartesian components of the various sublattices for each field direction are given in Table~\ref{tab:ansatze}.}
\label{fig:cartoon} 
\end{figure}

\vspace*{-0.3cm}
Following earlier works~\cite{Lee2015,Lee2016,Ducatman2018,Rousochatzakis2018}, we consider here the minimal microscopic $J$-$K$-$\Gamma$ model mentioned above, supplemented with a Zeeman term $\mc{H}_\text{Z}$ to describe the coupling to the external field ${\bf H}$. The total Hamiltonian then reads 
\sbe\label{eq:Hamiltonian0}
\mc{H}\!=\!\sum_t\sum_{\langle ij\rangle\in t}\mc{H}_{ij}^t+\mc{H}_\text{Z}\,,
\see 
where
\sbe\label{eq:Hamiltonian}
\begin{array}{c}
\mc{H}_{ij}^t=J\vec{S}_i\cdot\vec{S}_j+K S_i^{\alpha_t}S_j^{\alpha_t}+\sigma_t\Gamma(S_i^{\beta_t}S_j^{\gamma_t}+S_i^{\gamma_t}S_j^{\beta_t})\,, \\[1ex]
\mc{H}_\text{Z}=-\mu_B\vec{H}\cdot\sum_{i}\vec{g}_i\cdot \vec{S}_i \,.
\end{array}
\see
Here ${\bf S}_i$ denotes the pseudo-spin $j_{\text{eff}}\!=\!1/2$ operator at site $i$, $t\in\{x,y,z,x',y'\}$ labels the five different types of NN Ir-Ir bonds and $(\alpha_t,\beta_t,\gamma_t)\!=\!(x,y,z)$, $(y,z,x)$, and $(z,x,y)$ for $t\in\{x,x'\}$, $\{y,y'\}$, and $\{z\}$, respectively. The prefactor $\sigma_t$ equals $+1$ for $t\in\{x,y',z\}$ and $-1$ for $t\in\{y,x'\}$, see $\pm$ symbols in Figs.~\ref{fig:lattice} and \ref{fig:cartoon}. This overall sign structure of the $\Gamma$ interactions  derives from the symmetries mentioned above~\cite{Lee2015} and our choice of the $xyz$-frame in Eq.~(\ref{eq:xyzframe}). 
Finally, $\vec{g}_i$ stands for the ${\bf g}$-tensor of the $i$-th Ir ion. As discussed by Ruiz {\it et al}~\cite{Ruiz2017}, these tensors carry a site-dependent, staggered off-diagonal element $g_{ab}$. Specifically, in the orthorhombic frame,  
\sbe\label{eq:gtensor}
\vec{g}_{i}={\bf g}_{\rm\small diag}  + p_i {\bf g}_{\rm\small off-diag}  \equiv 
\begin{pmatrix}
g_{aa} & 0 & 0\\
0 & g_{bb} & 0\\
0 & 0 & g_{cc}\\
\end{pmatrix}
+p_i 
\begin{pmatrix}
0 & g_{ab} & 0\\
g_{ab} & 0 & 0\\
0 & 0 & 0\\
\end{pmatrix},
\see
where $p_i=+1$ for spins on the $xy$ chains and $-1$ for spins on the $x'y'$ chains. Here we take $g_{aa}\!=\!g_{bb}\!=\!g_{cc}\!=\!2$ and $g_{ab}\!=\!0.1$. 
We note that, depending on the direction of the field, some of the discrete symmetries mentioned above may or may not be preserved, see Table~\ref{tab:symmetries}. A field along $\bf a$, for example, breaks both $C_{2\bf b}$ and $C_{2\bf c}$, but still respects $C_{2\bf a}$, $\Theta C_{2\bf b}$ and $\Theta C_{2\bf c}$, where $\Theta$ is the time reversal operation.  

In the following we restrict ourselves to the so-called $K$-region of the parameter space with dominant Kitaev interaction, which is believed to be relevant for $\beta$-Li$_2$IrO$_3$~\cite{Ducatman2018}, and fix the parameters to  the representative set given in Eq.~(\ref{eq:JKGammaPars}). 

\vspace*{-0.3cm}
\section{Unified description of $\beta$-Li$_2$IrO$_3$ for ${\bf H}$ along ${{\bf a}}$, ${{\bf b}}$ and ${{\bf c}}$ axes}\label{sec:ansatze}
\vspace*{-0.3cm}
\subsection{General spin sublattice structure}\label{sec:SpinSublattices}
\vspace*{-0.3cm}
The behavior of $\beta$-Li$_2$IrO$_3$ under a magnetic field along the three orthorhombic directions can be described in a unified manner as shown in Fig.~\ref{fig:cartoon}. For all three directions, ${\bf a}$, ${\bf b}$ and ${\bf c}$, the system goes through a low-field phase ($0\!\le\!H\!<\!H^\ast$) with six spin sublattices [${\bf A}$, ${\bf B}$ and ${\bf C}$ along the $xy$-chains, and ${\bf A}'$, ${\bf B}'$ and ${\bf C}'$ along the $x'y'$-chains, see Fig.~\ref{fig:cartoon}\,(a)], followed by a high-field canted phase ($H^\ast\!<\!H\!<\!H^{\ast\ast}$) with two spin sublattices [${\bf F}$ along the $xy$-chains and ${\bf F}'$ along the $x'y'$ chains, see Fig.~\ref{fig:cartoon}\,(b)]. 
The high-field phase terminates at $H^{\ast\ast}\!=\!\infty$ for {\Ha} and {\Hb} (with a small zigzag component remaining if $g_{ab}\!\neq\!0$, see Appendix~\ref{app:ansatze}), whereas $H_{\bf c}^{\ast\ast}$ is finite and the classical state reached at $H_{\bf c}^{\ast\ast}$ is the fully polarized state.

\begin{figure}[!t]
{\includegraphics[width=\columnwidth]{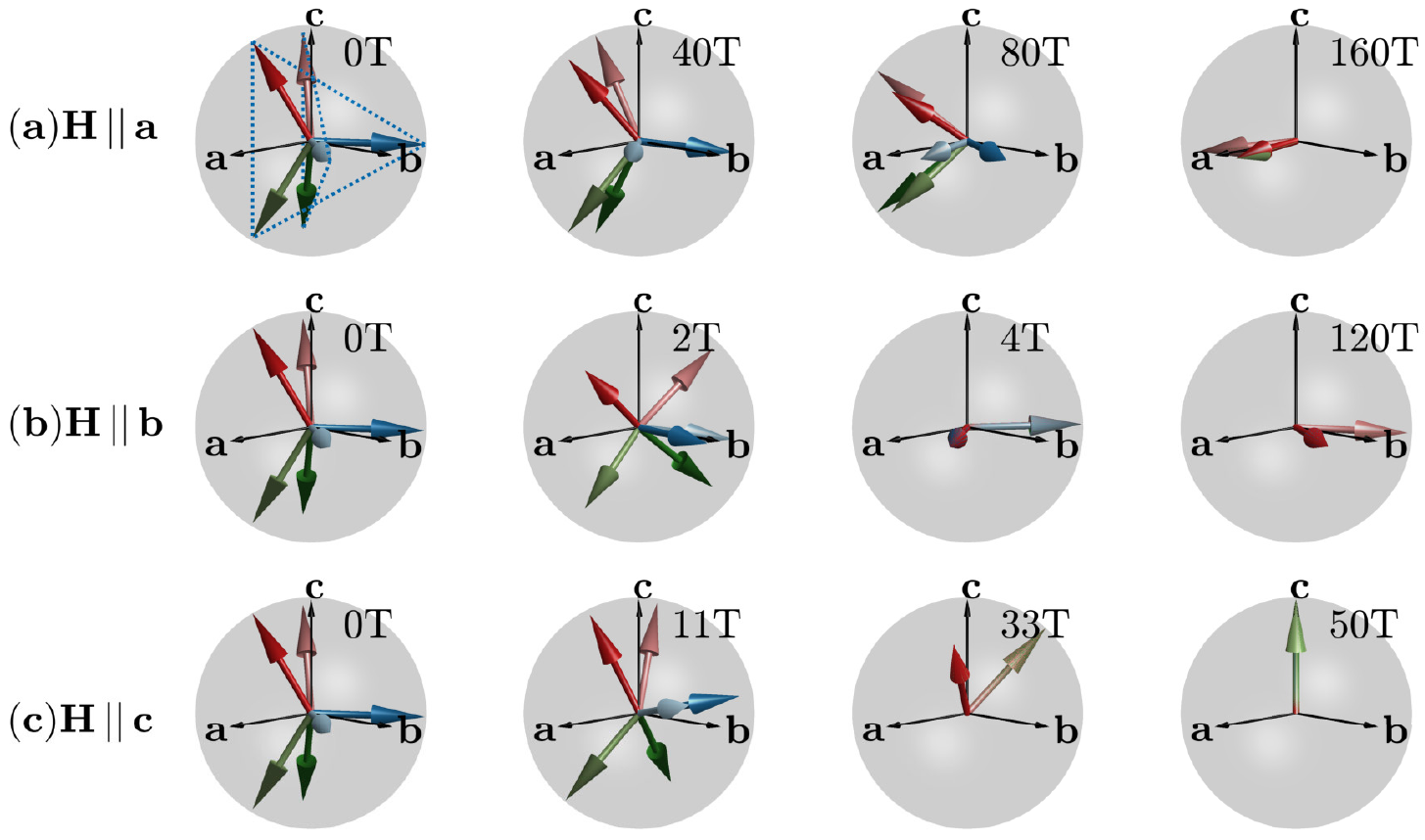}}
\caption{Snapshots of representative spin configurations for ${\bf H}$ along $\bf a$ (first row), $\bf b$ (second row), or $\bf c$ (third row). Each color represents one of the six sublattices of the zero-field state. The dashed lines in the upper left panel depict the nearly coplanar, 120$^\circ$ order of the ${\bf A B C}$ and ${\bf A}'{\bf B}'{\bf C}'$ sublattices~\cite{Ducatman2018}.}\label{fig:spinConfig}
\end{figure}

Figure~\ref{fig:spinConfig} shows a series of representative snapshots, of various ground state configurations for different field directions and strengths  (obtained from numerical minimization of the classical ans\"atze discussed below). As discussed in Ref.~[\onlinecite{Ducatman2018}], in the zero-field state, the three sublattices ${\bf A}$, ${\bf B}$ and ${\bf C}$ along the $xy$ chains form a nearly coplanar 120$^\circ$ state, and the three sublattices ${\bf A}'$, ${\bf B}'$ and ${\bf C}'$ along the $x'y'$ chains form another such nearly 120$^\circ$ structure, on a different plane, see dotted blue triangles at the top left panel of Fig.~\ref{fig:spinConfig}. Under a magnetic field, the three sublattices of each given chain cant toward each other and eventually get aligned at the characteristic field $H^\ast$ where ${\bf A}\!=\!{\bf B}\!=\!{\bf C}\!\equiv\!{\bf F}$ and ${\bf A}'\!=\!{\bf B}'\!=\!{\bf C}'\!\equiv\!{\bf F}'$. For fields along ${\bf a}$ and ${\bf b}$, this intra-chain alignment happens continuously, whereas for fields along ${\bf c}$ it happens abruptly. Above $H^\ast$, ${\bf F}$ and ${\bf F}'$ cant toward the field in a non-uniform way and at a pace that is strongly dependent on the field direction.

\vspace*{-0.3cm}
\subsection{Basic characterization of the low-field phase ($H\!<\!H^\ast$)}\label{sec:parametrization1}
\begin{figure*}[!t]
{\includegraphics[width=\textwidth]{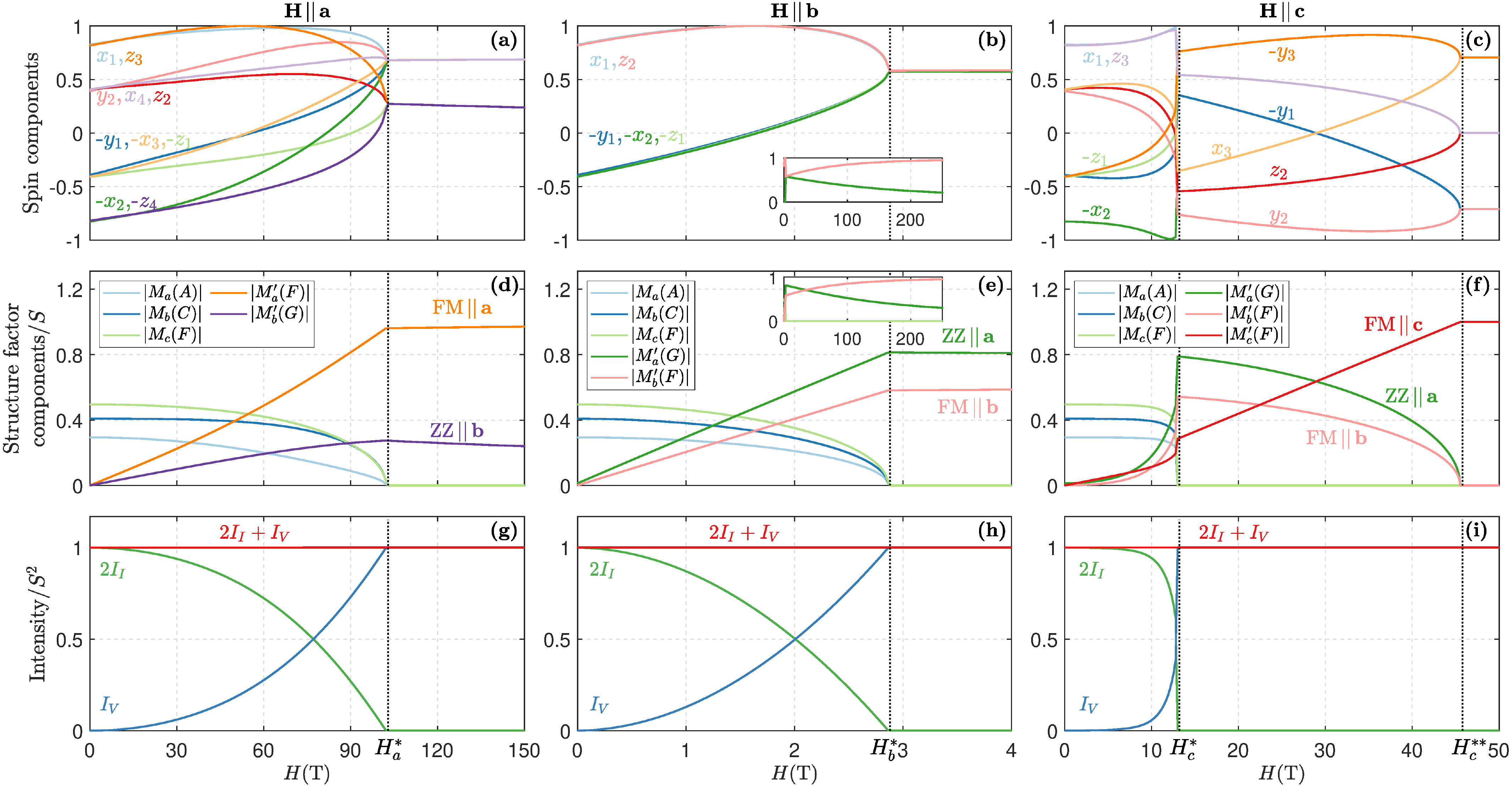}}
\caption{Field evolution of Cartesian spin components (a-c), dominant symmetry-resolved static structure factors (d-f) [see Fig.~\ref{fig:SmallComponents} for the remaining much weaker components], and Bragg peak intensities $I_I$, $I_{V}$ and $I_{\text{tot}}$ (g-i). The insets in (b) and (e) show the high-field behavior.}\label{fig:AnsatzResults} 
\end{figure*}

\begin{figure}[!b]
\includegraphics[width=0.7\columnwidth]{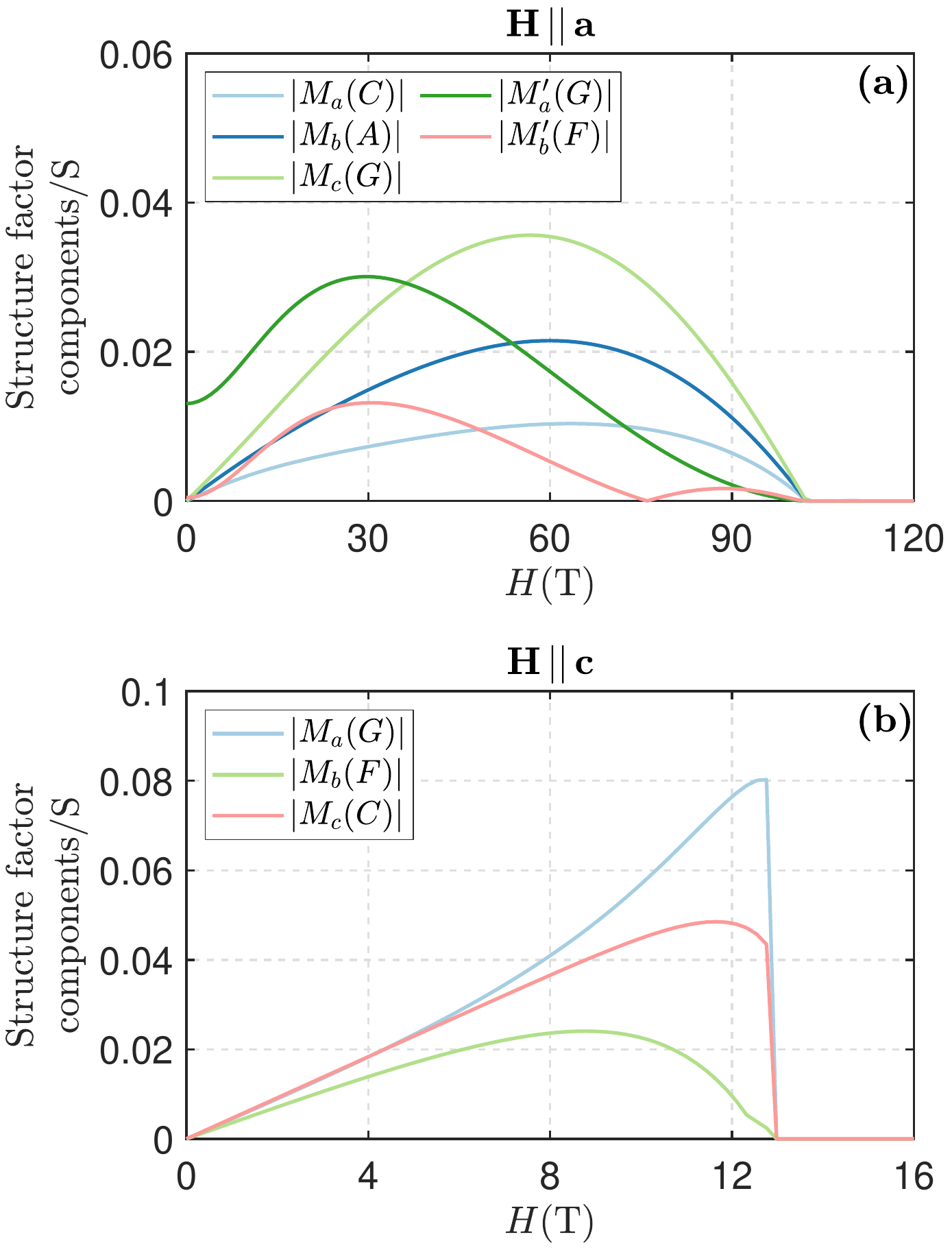}
\caption{Field dependence of the structure factors $M_a(C)$, $M_b(A)$, $M_c(G)$, $M'_a(G)$ and $M'_b(F)$ generated for ${\bf H}$ along ${\bf a}$ (a), and $M_a(G)$, $M_b(F)$ and $M_c(C)$ generated along ${\bf c}$ (b).}\label{fig:SmallComponents} 
\end{figure}

\vspace*{-0.3cm}
The individual Cartesian spin components of the various configurations are related to each other in a specific way, see the parametrization in Table~\ref{tab:ansatze}. For each given spin sublattice, a spin length constraint must be imposed, for example $x_1^2+y_1^2+z_1^2\!=\!1$ for the ${\bf A}$ sublattice, $2x_3^2+z_3^2\!=\!1$ for the ${\bf C}$ sublattice of the {\Ha} case, etc.  
The field dependence of the Cartesian components can be obtained by a numerical minimization of the total energy of the system [see Eqs.~(\ref{eq:energy_b1}), (\ref{eq:energy_a1}) and (\ref{eq:energy_c1})], and the results are shown in Figs.~\ref{fig:AnsatzResults}\,(a-c) as a function of the field. 
Equivalently, the spin configurations can be described in terms of the associated symmetry-resolved static structure factors, and the same is true for the total energy [see Eqs.~(\ref{eq:energy_b2}), (\ref{eq:energy_a2}) and (\ref{eq:energy_c2})]. The structure factors obey the same number of constraints as the Cartesian components [the relations between the two are given in Eqs.~(\ref{eq:SFvsCartesianb}), (\ref{eq:SFvsCartesiana}) and (\ref{eq:SFvsCartesianc})], and their evolution with field are shown in Figs.~\ref{fig:AnsatzResults}\,(d-f) and \ref{fig:SmallComponents}.

\begin{table}[!b]
\renewcommand{\arraystretch}{1.3}
\centering
\begin{tabular}{L{0.5cm} L{0.5cm} C{2.2cm} C{2.2cm} C{2.2cm}}
\toprule[1.pt]
&& {\Ha} & {\Hb}  & {\Hc} \\ 
\midrule[1.pt]
&${\bf A}$ & $S [x_1, y_1, z_1]$ &  $S [x_1, y_1, z_1]$ &  $S [x_1, y_1, z_1]$ \\
&${\bf A}'$ & $S [y_2, x_2, z_2]$ &  $S [y_1, x_1, z_1]$ &  $S [y_1, x_1, z_1]$ \\
&${\bf B}$ & $S [-y_1, -x_1, z_1]$ &  $S [-y_1, -x_1, z_1]$ &  $S [-y_2, -x_2, z_2]$ \\
&${\bf B}'$ & $S [-x_2, -y_2, z_2]$ &  $S [-x_1, -y_1, z_1]$ &  $S [-x_2, -y_2, z_2]$ \\
\rot{\rlap{~~$0\!\le\!H\!<\!H^\ast$}}&${\bf C}$ & $S [-x_3, x_3, -z_3]$ &  $S [-x_2, x_2, -z_2]$ &  $S [-y_3, x_3, -z_3]$ \\
&${\bf C}'$ & $S [x_4, -x_4, -z_4]$ &  $S [x_2, -x_2, -z_2]$ &  $S [x_3, -y_3, -z_3]$ \\
\midrule[1.pt]
&${\bf F}$ & $S [x_1, -x_1, z_1]$ &  $S [x_1, -x_1, z_1]$ &  $S [x_1, y_1, z_1]$ \\
\rot{\rlap{\!$ H\ge\!H^{\ast}$}}&${\bf F}'$ & $S [x_1, -x_1, -z_1]$ &  $S [-x_1, x_1, z_1]$ &  $S [y_1, x_1, z_1]$\\
\bottomrule[1.pt]
\end{tabular}
\setlength{\tabcolsep}{3em}
\caption{Cartesian components of the spin sublattices of the three ans\"atze analyzed here (see Appendix~\ref{app:ansatze} for more details). There are six spin sublattices for $0\!\le\!H\!<\!H^\ast$ (${\bf A}$, ${\bf B}$ and ${\bf C}$ along the $xy$-chains and ${\bf A}'$, ${\bf B}'$ and ${\bf C}'$ along the $x'y'$-chains), and two sublattices for $H\!>\!H^\ast$ (${\bf F}$ along the $xy$-chains and ${\bf F}'$ along the $x'y'$-chains), see Fig.~\ref{fig:cartoon}.} 
\label{tab:ansatze}
\renewcommand{\arraystretch}{1}
\end{table}

The low-field phase for {\Hb} is described by five Cartesian components ($x_1$, $y_1$, $z_1$, $x_2$ and $z_2$) or, equivalently, by five structure factors~\cite{Rousochatzakis2018}: Three modulated ${\bf Q}\!=\!\frac{2}{3}\hat{{\bf a}}$ components $M_a(A)$, $M_b(C)$ and $M_c(F)$, and two uniform ${\bf Q}\!=\!0$ components $M'_a(G)$ and $M'_b(F)$. 
This precise combination is in fact present for all three orthorhombic directions for $0\!\le\!H\!<\!H^\ast$, as it is a property of the zero-field state. 
In particular, the uniform components $M_a'(G)$ and $M_b'(F)$ of the zero-field order reflect the deviation from the perfect 120$^\circ$ coplanar order mentioned above, see detailed analysis in Ref.~[\onlinecite{Ducatman2018}].
Note further that the modulated components $M_a(A)$, $M_b(C)$ and $M_c(F)$ belong to the $\Gamma_4$ irreducible representation, in agreement with experiment~\cite{Biffin2014a}. 
The five structure factors satisfy two constrains which, when normalized appropriately [see Appendix~\ref{app:SSFconventions}], can be combined to give the Bragg peak intensity sum rule observed experimentally~\cite{Ruiz2017}. Namely, 
\sbe\label{eq:ISR0}
\!\!I_{\text{tot}} = 2 I_I + I_V = S^2\,,
\see
where
\sbe\label{eq:ISRb}
\renewcommand{\arraystretch}{1.2}
\begin{array}{c}
\!I_I=\!\!|M_a(A)|^2\!+\!|M_b(C)|^2\!+\!|M_c(F)|^2\!\equiv\!I_{I,\Gamma_4}\,,\\
\!I_V=|M'_a(G)|^2\!+\!|M'_b(F)|^2\,.
\end{array}
\see

Turning to the low-field phase for {\Ha}, here we have ten Cartesian components (see Table~\ref{tab:ansatze}) or, equivalently, ten structure factors: six modulated components (the three zero-field components plus three induced by the field) and four uniform components (the two zero-field components plus two induced by the field): 
\sbe
\renewcommand{\arraystretch}{1.2}
\!\!\!\!\begin{array}{ll}
H\!\geq\!0: & M_a(A),~M_b(C),~M_c(F)~\text{and}~M'_a(G),~M'_b(F)\,,\\
H\!>\!0: & M_a(C),~M_b(A),~M_c(G)~\text{and}~M'_a(F),~M'_b(G)\,.
\end{array}
\see 
Hence, a field along ${\bf a}$ induces a finite FM component $M_a'(F)$ (which couples explicitly to the Zeeman field), and a finite zigzag component $M_b'(G)$ along ${\bf b}$. The latter can become relatively large with field [see Fig.~\ref{fig:AnsatzResults}\,(d)] and should be observable experimentally, unlike the components $M'_a(G)$ and $M'_b(F)$ which remain at least one order of magnitude smaller, see Fig.~\ref{fig:SmallComponents}\,(a).
The same is true for the field-induced modulated components $M_a(C)$, $M_b(A)$ and $M_c(G)$, which belong to the irreducible representation $\Gamma_2$ (see Table II of Ref.~[\onlinecite{Biffin2014a}]). 
Altogether, the ten structure factors satisfy four constraints, and one combination of them gives the Bragg peak intensity sum rule of Eq.~(\ref{eq:ISR0}), where now 
\sbe\label{eq:ISRa}
\renewcommand{\arraystretch}{1.2}
\!\!\!\!\begin{array}{c}
I_{I}=I_{I,\Gamma_4}\!+\!I_{I,\Gamma_2} \simeq I_{I,\Gamma_4}\,,\\
I_{I,\Gamma_2}=|M_a(C)|^2\!+\!|M_b(A)|^2\!+\!|M_c(G)|^2 \ll I_{I,\Gamma_4}\,,\\
I_V=|M'_a(G)|^2\!+\!|M'_b(F)|^2\!+\!|M'_a(F)|^2\!+\!|M'_b(G)|^2\,.
\end{array}
\see

The low-field phase for {\Hc} is described by nine Cartesian components (see Table~\ref{tab:ansatze}) or by nine structure factors: 
\sbe
\renewcommand{\arraystretch}{1.2}
\!\!\!\!\!\begin{array}{ll}
H\!\geq\!0: & M_a(A),~M_b(C),~M_c(F)~\text{and}~M'_a(G),~M'_b(F)\,,\\
H\!>\!0: & M_a(G),~M_b(F),~M_c(C)~\text{and}~M'_c(F)\,.
\end{array}
\see 
Here the field induces three modulated components ($M_a(G)$, $M_b(F)$ and $M_c(C)$) and one uniform component $M'_c(F)$ (which couples directly to the Zeeman field). The modulated components belong to the irreducible representation $\Gamma_3$ (see Table II of Ref.~[\onlinecite{Biffin2014a}]), and, as it turns out, they remain at least one order of magnitude smaller than the dominant $\Gamma_4$ components, see Fig.~\ref{fig:SmallComponents}\,(b). 
Altogether, the nine structure factors satisfy three constraints, and one combination of them gives the Bragg peak intensity sum rule of Eq.~(\ref{eq:ISR0}), where now 
\sbe\label{eq:ISRc}
\renewcommand{\arraystretch}{1.2}
\!\!\!\!\begin{array}{c}
I_{I}=I_{I,\Gamma_4}\!+\!I_{I,\Gamma_3} \simeq I_{I,\Gamma_4}\,, \\
I_{I,\Gamma_3}=|M_a(G)|^2\!+\!|M_b(F)|^2\!+\!|M_c(C)|^2 \ll I_{I,\Gamma_4}\,,\\
I_V=|M'_a(G)|^2\!+\!|M'_b(F)|^2\!+\!|M'_c(F)|^2\,.
\end{array}
\see
Let us emphasize that the fulfilment of the intensity sum rule Eq.~(\ref{eq:ISR0}) for all field directions and strengths is a direct fingerprint of the local spin length constraints. The numerical prefactor of $2$ in the definition $I_{\text{tot}}\!=\!2I_I\!+\!I_V$ reflects the fact that there are twice as many Bragg peaks characterizing the modulated order (${\bf Q}\!=\!\pm\frac{2}{3}\hat{{\bf a}}$) compared to the peaks characterizing the uniform order (${\bf Q}\!=\!0$), see detailed analysis and a general proof of Eq.~(\ref{eq:ISR0}) in Appendix~\ref{app:Constraints}.

Note finally that some of the uniform components generated for ${\bf H}$ along ${\bf a}$ and ${\bf c}$ give rise to a finite magnetic torque signal, which will be examined separately in Sec.~\ref{sec:Torque}. 

\begin{figure*}[!t]
{\includegraphics[width=0.85\linewidth]{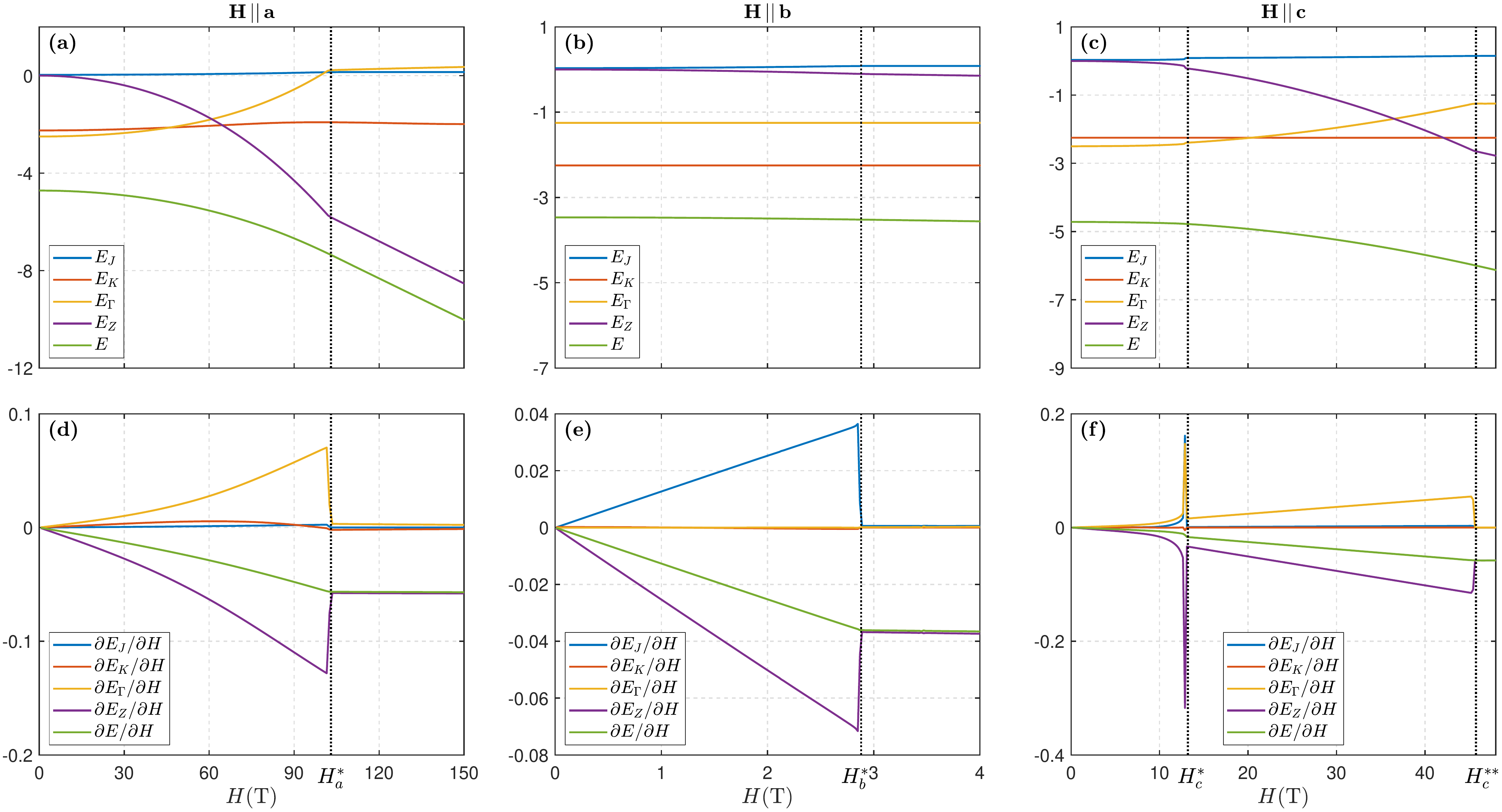}}
\caption{(a-c) Evolution of the various contributions to the energy ($E_J$, $E_K$, $E_\Gamma$ and $E_H$ denote the contributions from $J$, $K$, $\Gamma$ and the Zeeman field, respectively) with $H$, for ${\bf H}$ along ${\bf a}$,  ${\bf b}$ and  ${\bf c}$.  (d-f): Evolution of the first derivates of $E_J$, $E_K$, $E_\Gamma$ and $E_H$ with respect to the field $H$. All  energies are given  in meV.}\label{fig:Energy}
\end{figure*}

\vspace*{-0.3cm}
\subsection{Basic characterization of the high-field phase ($H^{\ast}\!<\!H\!<\!H^{\ast\ast}$)}\label{sec:parametrization2}
\vspace*{-0.3cm}
For $H\!>\!H^\ast$, all modulated components vanish identically, and we are left with uniform structure factors only. In particular, for ${\bf H}$ along ${\bf a}$ and ${\bf b}$, there are only two uniform components, a FM component along the field and a zigzag component perpendicular to the field.  For {\Hc}, there is an additional FM component $M'_b(F)$ perpendicular to the field. In terms of the two spin sublattices ${\bf F}$ and ${\bf F}'$ of Fig.~\ref{fig:cartoon}\,(b), the FM component is proportional to ${\bf F}+{\bf F}'$ and the zigzag component is proportional to ${\bf F}-{\bf F}'$. 
The direction of the zigzag component depends on the direction of the field. When {\Ha}, ${\bf F}-{\bf F}' \!=\! 2S z_1 \hat{{\bf z}}$, see Table~\ref{tab:ansatze}, and therefore the zigzag component is fixed along ${\bf b}$. 
By contrast, the zigzag component is fixed along ${\bf a}$ when ${\bf H}$ points along ${\bf b}$ or ${\bf c}$, with ${\bf F}\!-\!{\bf F}'\!=\!2\sqrt{2}S x_1\hat{{\bf a}}$ and ${\bf F}\!-\!{\bf F}'\!=\!\sqrt{2}S(x_1\!-\!y_1)\hat{\bf a}$,  respectively, see Table~\ref{tab:ansatze}. Note also that, for $H\!\ge\!H^\ast$, the spins lie on the $ab$-plane for {\Ha} and {\Hb}, but for {\Hc} the spin plane changes continuously. This is related to the fact that the uniform components of the zero-field state all lie in the $ab$-plane, and so a field applied in this plane will merely reorganize these components and not rotate them out of the plane, unlike what happens for {\Hc}.

The zigzag component disappears at a characteristic field $H^{\ast\ast}$. As mentioned above, $H^{\ast\ast}$ is infinite for ${\bf H}$ along ${\bf a}$ and ${\bf b}$ but finite for {\Hc}, with (see Appendix~\ref{app:Halongc}) 
\sbe\label{eq:Hcstarstar}
\mu_B H_{\bf c}^{\ast\ast}
=\Big( \Gamma+2J+\sqrt{(\Gamma-2J)^2+8\Gamma^2} ~\Big)\frac{S}{2g_{cc}} \,,
\see 
which, for $J\!\ll\!|\Gamma|$, reduces to 
\sbe\label{eq:Hcstarstar2}
\mu_B H_{\bf c}^{\ast\ast}\!\simeq\!\Big(\frac{4}{3}J\!+\!|\Gamma| \Big)\frac{S}{g_{cc}} \,.
\see 
According to this relation, $H_{\bf c}^{\ast\ast}$ depends mostly on $\Gamma$, with $H_{\bf c}^{\ast\ast}\!\sim\!45$~T for the coupling parameters of Eq.~(\ref{eq:JKGammaPars}).

\vspace*{-0.3cm}
\subsection{Robustness of high-field zigzag orders}\label{sec:Robustness}
\vspace*{-0.3cm}
We now discuss why the various high-field zigzag orders remain robust up to very high fields, for all three orthorhombic directions. The most direct way to see this is to express the total energies $E_{\bf b}$, $E_{\bf a}$ and $E_{\bf c}$ in terms of the various static structure factors, see Eqs.~(\ref{eq:energy_b3}), (\ref{eq:energy_a3}) and (\ref{eq:energy_c3}), respectively. It turns out that  $E_{\bf b}$ and $E_{\bf c}$ contain an explicit cross-coupling term between $M'_a(G)$ and $M'_b(F)$, 
\sbe\label{eq:cross-coupling1}
-\sqrt{2}\Gamma M'_a(G) M'_b(F)\,,
\see
while $E_{\bf a}$ contains an explicit cross-coupling term between $M'_a(F)$ and $M'_b(G)$, 
\sbe\label{eq:cross-coupling2}
-\sqrt{2}\Gamma M'_a(F) M'_b(G)\,.
\see
The presence of these terms reveal that the qualitative reason why it is energetically favorable for the system to sustain appreciable zigzag orders up to high fields is the strong $\Gamma$ interaction.
Of course, the actual quantitative details for each field direction derive from the minimization of the total energies under the given constraints. For example, the analytical expression Eq.~(\ref{eq:Hcstarstar}) for $H_{\bf c}^{\ast\ast}$ can be derived by minimizing $E_{\bf c}$ in Eq.~(\ref{eq:energy_c3}) under the single constraint $|M'_a(G)|^2\!+\!|M'_b(F)|^2\!+\!|M'_c(F)|^2\!=\!S^2$.

\vspace*{-0.3cm}
\subsection{Dependence of $H^\ast$ on microscopic coupling parameters}\label{sec:Hstar}
\vspace*{-0.3cm}
The characteristic field $H^\ast$ marks the disappearance of the modulated components (and $M_a'(G)$ and $M_b'(F)$ for {\Ha}). As mentioned earlier, this transition is continuous for {\Ha} and {\Hb}, but of first-order for {\Hc}, see Figs.~\ref{fig:AnsatzResults}\,(a- f). 
Furthermore, the value of $H^\ast$ depends strongly on the direction of the field. For the coupling parameters of Eq.~(\ref{eq:JKGammaPars}), $H_{\bf a}^\ast\!\sim\!102$~T, $H_{\bf b}^\ast\sim2.88$~T and $H_{\bf c}^\ast\!\sim\!13$~T. This large difference between the critical fields along different directions is related to the strongly anisotropic character of the Hamiltonian, and the different role of the various couplings in each case. For example, as we discussed in Ref.~[\onlinecite{Rousochatzakis2018}], in the parameter regime of interest, $H_{\bf b}^\ast$ depends only on $J$, which is why $H_{\bf b}^\ast$ is very small. 

We will now show that $H_{\bf a}^\ast$ and $H_{\bf c}^\ast$ do not depend on $K$ but only on $J$ and $\Gamma$, and that the inequality $J\ll |\Gamma|$ explains why these critical fields are larger compared to $H_{\bf b}^\ast$. 
To this end, we will vary the parameters of the model and take a closer look at the evolution of the various contributions to the total energy with the field. Figure~\ref{fig:Energy}\,(a-c) shows the field-driven evolution of $E_J$, $E_K$, $E_\Gamma$ and $E_\text{Z}$, which denote the contributions from $J$, $K$ and $\Gamma$ interactions and the Zeeman energy, respectively. 
The corresponding derivatives of these energies with respect to $H$ are shown in Fig.~\ref{fig:Energy}\,(d-f). The main finding is that, in the parameter regime of interest, $E_K$ remains almost insensitive to $H$, and this is true for all field directions. This means that the Zeeman field does not act against $K$, which explains why none of the critical fields $H^\ast$ depends on the dominant coupling of the theory. 
The results also show that, unlike $H_{\bf a}^\ast$ and $H_{\bf c}^\ast$, the critical field $H_{\bf b}^\ast$ depends only on $J$ and not on $\Gamma$; this is the consequence  of the fact that $E_\Gamma$ does not change with $H$ in this direction.
These arguments can be formulated mathematically by the following relations that arise from a classical version of Feynman-Hellmann theorem (see Appendix~\ref{app:FeynmanHellmann}):
\sbe\label{eq:FeynmanHellmann}
\begin{array}{c}
\mc{N} \frac{\partial}{\partial J} m_\parallel(J,K,\Gamma,H) = -\frac{\partial}{\partial H} E_J(J,K,\Gamma,H)/J \,, \\[1.5ex]
\mc{N} \frac{\partial}{\partial K} m_\parallel(J,K,\Gamma,H) = -\frac{\partial}{\partial H} E_K(J,K,\Gamma,H)/K \,, \\[1.5ex]
\mc{N} \frac{\partial}{\partial J} m_\parallel(J,K,\Gamma,H) = - \frac{\partial}{\partial H} E_\Gamma(J,K,\Gamma,H)/K \,, 
\end{array}
\see
where $\mc{N} $ is the total number of spins and $m_\parallel(J,K,\Gamma,H)$ is the magnetization per site along the field.  According to these relations, the fact that $\partial E_K/\partial H\!\approx\!0$ implies that $\partial m_\parallel/\partial K\!\approx\!0$, i.e., that the whole magnetization process does not depend on $K$. Likewise, the fact that $\partial E_\Gamma/\partial H\!\approx\!0$ for {\Hb} implies that $\partial m_\parallel/\partial \Gamma\!\approx\!0$, and therefore the whole magnetization process depends only on $J$ in this field direction.

\begin{figure*}[!t]
{\includegraphics[width=0.95\textwidth]{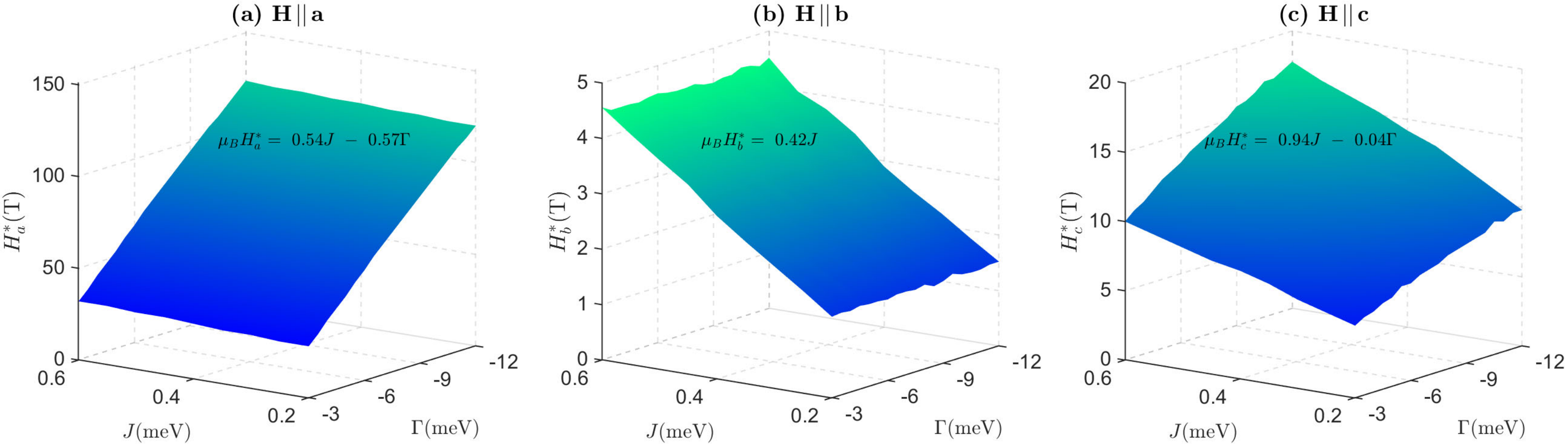}}
\caption{Variation of critical fields $H^\ast$ in the $J$-$\Gamma$ plane around the relevant parameter regime for $\beta$-Li$_2$IrO$_3$.}\label{fig:Hstar}
\end{figure*}

We can go one step further and extract the actual dependence of the critical fields on the relevant couplings by computing these fields for a wider range of parameters. The results are shown in Fig.~\ref{fig:Hstar} and demonstrate that the critical fields $H_{\bf a}^\ast$ and $H_{\bf c}^\ast$ depend almost perfectly linearly on $J$ and $\Gamma$. Fitting the numerical data for $H^\ast$ gives, in particular,
\sbe\label{eq:Hacstar}
\begin{array}{l}
\mu_B H_{\bf a}^\ast \simeq  \Big(0.54 J + 0.57 |\Gamma| \Big) \frac{4S}{g_{aa}},\\[3pt]
\mu_B H_{\bf b}^\ast \simeq 0.42 J ~\left(\frac{4S}{g_{bb}}\right)\,,\\[3pt]
\mu_B H_{\bf c}^\ast \simeq \Big(0.94 J + 0.04 |\Gamma| \Big) \frac{4S}{g_{cc}}\,.
\end{array}
\see
Thus, besides the independence of $H_{\bf b}^\ast$ on $K$ and $\Gamma$,~\footnote{Note that the coefficient $0.42$ in the second relation is slightly different from the corresponding coefficient $0.46$ reported previously~\cite{Rousochatzakis2018}, which was obtained from fitting the numerical data up to larger values of $J$, where the assumption of $H_{\bf b}^\ast$ being independent on $K$ and $\Gamma$ breaks down.}, we find that $H_{\bf a}^\ast$ is controlled mainly by $\Gamma$ (given that $J\!\ll\!|\Gamma|$), whereas $H_{\bf c}^\ast$ is controlled by both $J$ and $\Gamma$. Note that the coefficients appearing in Eqs.~(\ref{eq:Hacstar}) correspond to the value $g_{ab}\!=\!0.1$ whose sign and magnitude is chosen arbitrarily here. However, the coefficients do not depend much on this choice. For example, for $g_{ab}\!=\!0$ we get $\mu_B H_{\bf a}^\ast\!\simeq\!\Big(0.54 J \!+\! 0.59 |\Gamma| \Big) \frac{4S}{g_{aa}}$, $\mu_B H_{\bf b}^\ast\!\simeq\!0.45 J ~\left(\frac{4S}{g_{bb}}\right)$, while $H_{\bf c}^\ast$ remains unchanged.

\vspace*{-0.3cm}
\subsection{Symmetries}\label{sec:Symmetries}
\vspace*{-0.3cm}
Table~\ref{tab:symmetries} shows the symmetry properties of the various field-induced configurations for different field directions. The primitive translations (denoted by $\mc{T}$) are broken spontaneously in the low-field phase ($0\!<\!H\!<H^\ast$) due to the modulating components of the order. This symmetry is restored above $H^\ast$ with the disappearance of these components. Furthermore, the low-field phases preserve the inversion symmetries $\mc{I}$ around the centers of the FM dimers ${\bf A}{\bf A}$, ${\bf B}{\bf B}$, ${\bf A}'{\bf A}'$ or ${\bf B}'{\bf B}'$ of Fig.~\ref{fig:cartoon}\,(a), while the high-field phases preserve the inversion centers on all $x$, $y$, $x'$ and $y'$ bonds.

\begin{table*}[!t]
\renewcommand{\arraystretch}{1.3}
\centering
\begin{tabular}{L{2.7cm} C{0.8cm}C{0.8cm}C{0.8cm}C{0.8cm}C{0.8cm} C{0.8cm}C{0.8cm}C{0.8cm}C{0.8cm}C{0.8cm} C{0.8cm}C{0.8cm}C{0.8cm}C{0.8cm}C{0.8cm}}
\toprule[1.pt]
\multicolumn{1}{l}{field direction} 
&
\multicolumn{5}{c}{{\Ha}}  &
\multicolumn{5}{c}{{\Hb}}  &
\multicolumn{5}{c}{{\Hc}}  \\ 
\cmidrule(lr){2-6}
\cmidrule(lr){7-11}
\cmidrule(lr){12-16}
\multicolumn{1}{l}{Hamiltonian $\mc{H}$} & 
\multicolumn{1}{c}{$\mc{T}$} &
\multicolumn{1}{c}{$\mc{I}$} &
\multicolumn{1}{c}{$C_{2\bf a}$} &
\multicolumn{1}{c}{$\Theta C_{2\bf b}$}&
\multicolumn{1}{c}{$\Theta C_{2\bf c}$}&
\multicolumn{1}{c}{$\mc{T}$} &
\multicolumn{1}{c}{$\mc{I}$} &
\multicolumn{1}{c}{$\Theta C_{2\bf a}$} &
\multicolumn{1}{c}{$C_{2\bf b}$}&
\multicolumn{1}{c}{$\Theta C_{2\bf c}$}&
\multicolumn{1}{c}{$\mc{T}$} &
\multicolumn{1}{c}{$\mc{I}$} &
\multicolumn{1}{c}{$\Theta C_{2\bf a}$} &
\multicolumn{1}{c}{$\Theta C_{2\bf b}$}&
\multicolumn{1}{c}{$C_{2\bf c}$}\\
\midrule[1.pt]
state at $0\!<\!H\!<\!H^\ast$ & ${\cred\times}$& $\surd$ & ${\cred\times}$&${\cred\times}$&$\surd$ & ${\cred\times}$& $\surd$ & $\surd$ & $\surd$ & $\surd$ &${\cred\times}$& $\surd$ & $\surd$&${\cred\times}$&${\cred\times}$ \\
state at $H^\ast\!<\!H\!<\!H^{\ast\ast}$ & $\surd$&$\surd$& $\surd$ &$\surd$ & $\surd$ & $\surd$ & $\surd$ & $\surd$ & $\surd$ & $\surd$& $\surd$ & $\surd$& $\surd$ &${\cred\times}$&${\cred\times}$\\
state at $H\!>\!H^{\ast\ast}$ & $\surd$&$\surd$& $\surd$ &$\surd$ & $\surd$ & $\surd$ & $\surd$ & $\surd$ & $\surd$ & $\surd$& $\surd$ & $\surd$& $\surd$ &$\surd$&$\surd$\\
\bottomrule[1.pt]
\end{tabular}
\setlength{\tabcolsep}{3em}
\caption{Discrete symmetries of the Hamiltonian (see Sec.~\ref{sec:LatticeSymmetry}) and the various states discussed in the text. $\mc{T}$ denotes the primitive translations of the crystal, $\Theta$ is time reversal, and $\mc{I}$ denotes the inversion centers of the ferromagnetic dimers for $0\!<\!H\!<H^\ast$, or any inversion center of the structure for $H\!>\!H^\ast$. Note that $H_{\bf a}^{\ast\ast}\!=\!\infty$ and $H_{\bf b}^{\ast\ast}\!=\!\infty$, whereas $H_{\bf c}^{\ast\ast}$ is finite.}
\label{tab:symmetries}
\renewcommand{\arraystretch}{1}
\end{table*}

Let us now turn to the $C_2$-rotation symmetries discussed in Sec.~\ref{sec:LatticeSymmetry} or their combinations with time reversal $\Theta$. For {\Hb}, the symmetries $\Theta C_{2{\bf a}}$, $C_{2{\bf b}}$ and $\Theta C_{2{\bf c}}$ of the model are all preserved in both the low- and the high-field phases, emphasizing once again the special role of the ${\bf b}$ axis~\cite{Biffin2014a, Ruiz2017,Rousochatzakis2018}. 

For {\Ha}, on the other hand, among the three symmetries $C_{2{\bf a}}$, $\Theta C_{2{\bf b}}$ and $\Theta C_{2{\bf c}}$, the first two are broken spontaneously in the low-field phase due to $M'_a(G)$ and $M'_b(F)$. This symmetry breaking is associated with the choice of the overall sign of $M'_a(G)$ and $M'_b(F)$. One can see this more directly from the cross-coupling term of Eq.~(\ref{eq:energy_a2}), according to which the relative signs of $M'_a(G)$ and $M'_b(F)$ are fixed by the sign of $\Gamma$, but one can still change both signs at the same time without changing the energy. 
Note that, while a similar cross-coupling term appears between $M'_a(F)$ and $M'_b(G)$, see Eq.~(\ref{eq:cross-coupling2}), the individual signs of these two components are fixed by the Zeeman field which couples directly to $M'_a(F)$, see Appendix~\ref{app:Halonga}. 
The symmetries $C_{2{\bf a}}$ and $\Theta C_{2{\bf b}}$ are restored at $H_{\bf a}^\ast$ with the disappearance of the $M'_a(G)$ and $M'_b(F)$.

The situation for {\Hc} has one qualitative difference (besides the abrupt transition at $H_{\bf c}^\ast$). Here, among the three symmetries $\Theta C_{2\bf a}$, $\Theta C_{2\bf b}$ and $C_{2\bf c}$ of the model, the last two are broken spontaneously in both the low- and the high-field phase, and only get restored at $H\!\geq\!H_{\bf c}^{\ast\ast}$. 
The symmetry breaking occurs again due to $M'_a(G)$ and $M'_b(F)$, which couple via Eq.~(\ref{eq:cross-coupling1}). As above then, $\Gamma$ fixes the relative signs of $M'_a(G)$ and $M'_b(F)$, but the overall choice of the global sign remains arbitrary.
Altogether, unlike what happens along ${\bf a}$, the transition at $H_{\bf c}^\ast$ does not restore all broken symmetries, and one thus expects a second thermal phase transition at high fields, even after the disappearance of the modulated order. This will be shown explicitly in Sec.~\ref{sec:FiniteT}.

For completeness, let us recall that the zero-field state breaks $C_{2{\bf a}}$ and $C_{2{\bf c}}$, but respects $\Theta C_{2{\bf a}}$, $\Theta C_{2{\bf c}}$ and $C_{2{\bf b}}$~\cite{Ducatman2018}.

\vspace*{-0.3cm}
\section{Magnetization process \& the effect of quantum fluctuations}\label{sec:QFs}
\begin{figure*}
{\includegraphics[width=0.9\linewidth]{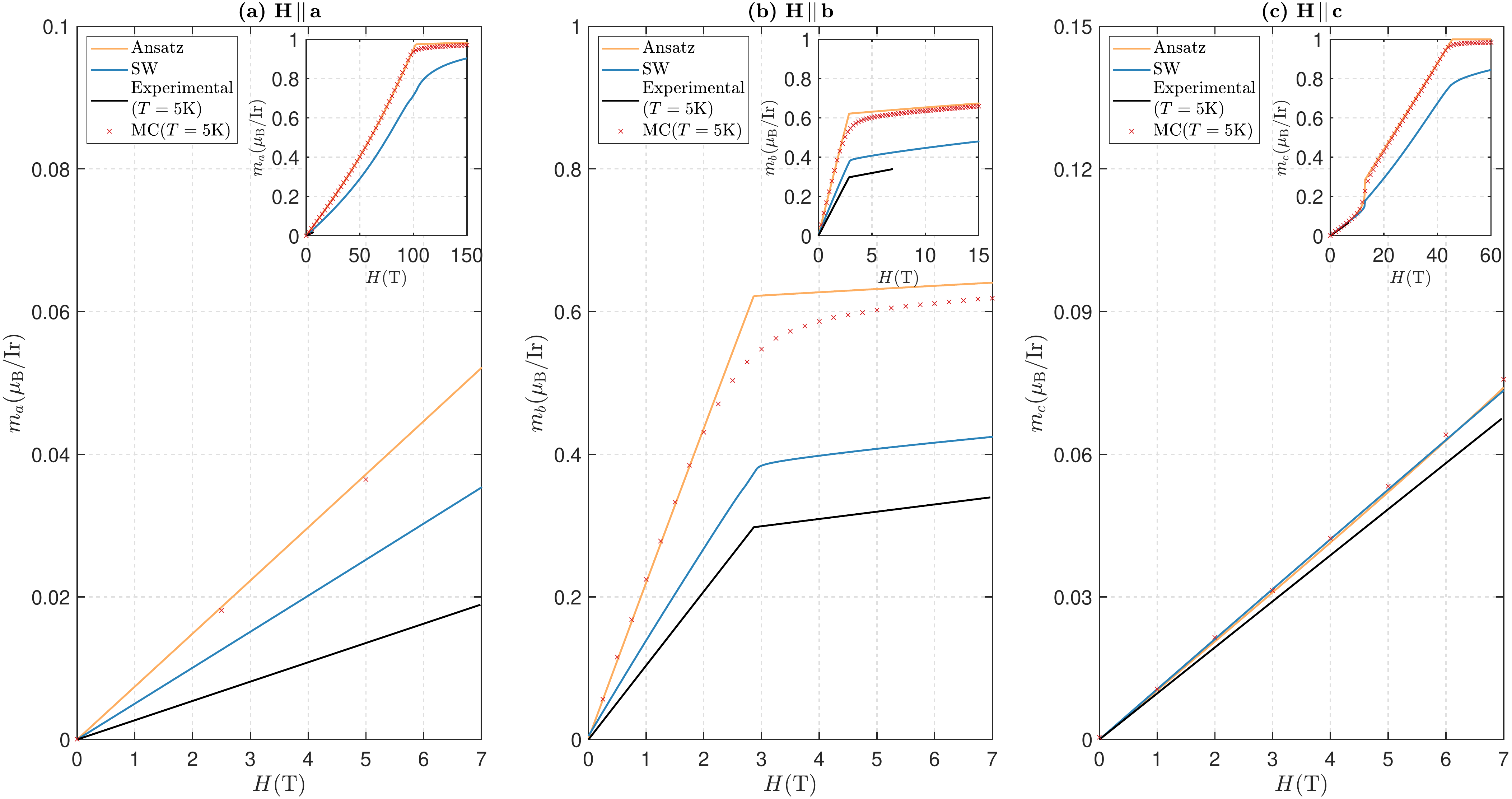}}
\caption{Main panels: Magnetization process up to $7$~T, obtained from the classical ans\"atze (orange line), the linear spin wave  approximation (blue line), and classical Monte Carlo simulations (red crosses). For comparison, we also show published experimental data (see supplementing material of Ref.~[\onlinecite{Ruiz2017}]). 
The insets at the upper-right corners show the computed magnetization curves up to much higher fields (up to $150$~T, $15$~T and $60$~T for panels (a), (b) and (c), respectively).}\label{fig:magnetization}
\end{figure*}

\vspace*{-0.3cm}
We now focus on the magnetization per site ${\bf m}$, defined as 
\sbe\label{eq:magn}
{\bf m} = \frac{1}{\mc{N}_{\text{m}}} \mu_B \left( {\bf g}_{\rm\small diag} \cdot \sum_\mu \langle {\bf S}_{\mu} \rangle + {\bf g}_{\rm\small off-diag} \cdot \sum_\mu p_\mu \langle {\bf S}_{\mu} \rangle \right)\,.
\see
Here $\mc{N}_{\text{m}}$ is the number of spins inside the magnetic unit cell ($\mc{N}_{\text{m}}\!=\!48$ for $H\!<\!H^\ast$ and $\mc{N}_{\text{m}}\!=\!2$ for $H\!>\!H^\ast$), $\mu\!=\!1$-$\mc{N}_{\text{m}}$, $\langle{\bf S}_{\mu}\rangle$ is the expectation value of the spin on the $\mu$-th sublattice, and ${\bf g}_{\rm\small diag}$, ${\bf g}_{\rm\small off-diag}$ and $p_\mu$ are defined in Eq.~(\ref{eq:gtensor}). 
Recalling that $p_\mu\!=\!+1$ for spins along the $xy$ chains and $-1$ for spins along the $x'y'$ chains [see Fig.~\ref{fig:lattice}], we see that the second contribution of Eq.~(\ref{eq:magn}) comes from the zigzag component of the order. This contribution vanishes for $g_{ab}\!=\!0$ and is about 5\% of the first term of Eq.~(\ref{eq:magn}) for $g_{ab}\!=\!0.1$.  
More explicitly, we have
\sbe\label{eq:magnetization}
\!\!\!\!{\bf m}\!=\!\!\left\{\!\!
\renewcommand{\arraystretch}{1.3}
\begin{array}{cl}
\Big[g_{aa} M'_a(F) \!+\! g_{ab} M'_b(G)\Big] \hat{{\bf a}}\!+\!\Big[g_{bb} M'_b(F) \!+\! g_{ab} M'_a(G)\Big] \hat{{\bf b}}, & \!\! {\bf H}\!\parallel\!{\bf a}\\
\Big[g_{bb} M'_b(F) \!+\! g_{ab} M'_a(G)\Big] \hat{{\bf b}}, & \!\!{\bf H}\!\parallel\!{\bf b}\\
g_{cc} M'_c(F) \hat{{\bf c}} \!+\!\Big[g_{bb} M'_b(F) \!+\! g_{ab} M'_a(G)\Big] \hat{{\bf b}}, & \!\!{\bf H}\!\parallel\!{\bf c}
\end{array}
\right.
\see
The magnetizations along the field $m_\parallel$ (denoted by $m_a$, $m_b$ and $m_c$ for ${\bf H}$ along ${\bf a}$, ${\bf b}$ and ${\bf c}$, respectively) are given by $m_a\!=\!g_{aa} M'_a(F)\!+\!g_{ab} M'_b(G)$, $m_b\!=\!g_{bb}M'_b(F)\!+\!g_{ab}M'_a(G)$, and $m_c\!=\!g_{cc}M'_c(F)$. Their evolutions with field are shown by the orange solid lines in Figs.~\ref{fig:magnetization}\,(a-c), and follow the general trend of $M'_a(F)$, $M'_b(F)$ and $M'_c(F)$, see Figs.~\ref{fig:AnsatzResults}\,(d-f).

In agreement with experiment, $m_b$ rises much faster than $m_a$ and $m_c$. 
 Furthermore, the magnetizations $m_a$ and $m_b$ first increase monotonously with the field, then show a kink at $H_{\bf a}^\ast$ and $H_{\bf b}^\ast$, respectively, and then increase at a much slower pace towards a limiting value that is determined by the ratios $g_{ab}/g_{aa}$ and $g_{ab}/g_{bb}$, respectively (see Appendix~\ref{app:ansatze}). 
By contrast, $m_c$ shows a finite jump (instead of a kink) at $H_{\bf c}^\ast$, reflecting the corresponding jumps in Fig.~\ref{fig:AnsatzResults}\,(f).  At higher fields, $m_c$ shows a kink at $H_{\bf c}^{**}$ and then saturates. Note that here the exact saturation is only true for classical spins, and the kink  in the classical magnetization will be smoothed by quantum fluctuations (as the spin Hamiltonian does not conserve rotations around the field axis, and the fully polarized state is not a true eigenstate).

Let us now compare these classical predictions for $m_\parallel$ with available experimental data published by Ruiz {\it et al} (see supplementing material in \cite{Ruiz2017}), which are shown in Fig.~\ref{fig:magnetization} by black lines. 
Quite generally, while the classical ans\"atze capture the observed magnetization processes qualitatively, there is a large quantitative discrepancy. For {\Hb}, for example, the classical prediction for the magnetization at $H_{\bf b}^\ast$ is about two times larger than the measured value. This deficiency has been recognized previously~\cite{Rousochatzakis2018}, and has led to the assertion that the system must feature strong quantum fluctuations due to the close proximity to the highly-frustrated $K$-$\Gamma$ line~\cite{Ducatman2018}.

Here we confirm  this hypothesis by calculating the leading $1/S$ corrections to the magnetization from quantum fluctuations. The details of this calculation are provided in Appendix~\ref{app:SW} and the renormalized magnetization curves are shown by the solid blue lines in Fig.~\ref{fig:magnetization}. The results show that already the leading $1/S$ corrections reduce the magnetization quite strongly, bringing the curves much closer to the measured data. While subleading higher-order corrections will reduce the magnetization even further, providing a better comparison  between theory and experiment, a final quantitative agreement will also require an appropriate re-adjustment of the microscopic couplings.

Importantly, our semiclassical results show further that, for {\Hc}, the magnitude of the magnetization jump at $H_{\bf c}^\ast$ is significantly reduced by quantum fluctuations, almost to the point that there is no visible change, including the overall slopes of the curves below and above $H_{\bf c}^\ast$. This renders the detection of this feature in magnetization measurements more challenging and probably explains the absence of the kink in recent measurements~\cite{Majumder2019}. The detection is even more challenging for powder samples  given that $m_c\!\ll\!m_b$. Nevertheless, as we will discuss next [Sec.~\ref{sec:Torque}], the transition at $H_{\bf c}^\ast$ should be still visible via the kink in the corresponding magnetic torque.

\begin{figure}[!b]
{\includegraphics[width=0.95\linewidth]{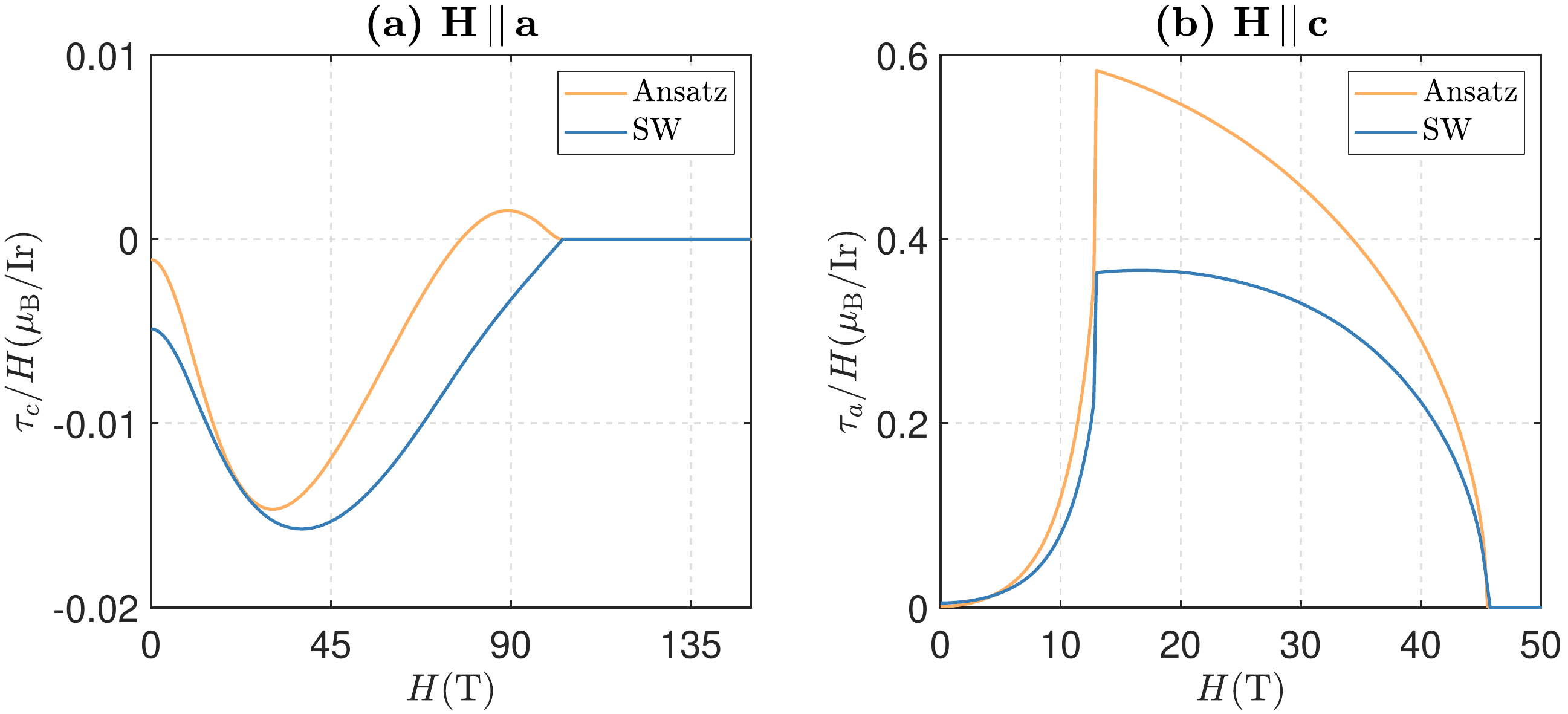}}
\caption{Field dependence of the torques computed with the classical ans\"atze (orange) and within the linear spin wave  approximation (blue).}
\label{fig:torque}
\end{figure}

\vspace*{-0.3cm}
\section{Magnetic torque}\label{sec:Torque}
\vspace*{-0.3cm}
According to Eq.~(\ref{eq:magnetization}), when {\Ha} and {\Hc}, the magnetization ${\bf m}$ develops a component perpendicular to ${\bf H}$. This implies the presence of a finite torque,  
\sbe\label{eq:Torque}
\renewcommand{\arraystretch}{1.2}
\begin{array}{c}
{\bf H}\!\parallel\!{\bf a}\!: \bs{\tau}= - \xi H ~\hat{{\bf c}},~~~ 
{\bf H}\!\parallel\!{\bf c}\!: \bs{\tau}= \xi H  ~\hat{{\bf a}},\\
\xi \equiv g_{bb} M'_b(F)+g_{ab}M'_a(G)\,.
\end{array}
\see
Interestingly, the expression for $\xi$ that gives the transverse components for {\Ha} and {\Hc} coincides with the expression for $m_b$, see Eq.~(\ref{eq:magnetization}).  
Note that, as we discussed in Sec.~\ref{sec:Symmetries}, the overall signs of $M'_a(G)$ and $M'_b(F)$ are chosen spontaneously by the system for both {\Ha} and {\Hc}, and therefore the sign of the torque (or $\xi$) is arbitrary for both directions. This aspect has further observable consequences, which will be discussed in Sec.~\ref{sec:Discussion}.

Figure~\ref{fig:torque}\,(b) shows the evolution of $\tau/H$ with $H$ for fields along ${\bf a}$ and ${\bf c}$, with and without harmonic spin-wave corrections. First of all, the torque for {\Ha} is about 40 times weaker than the torque for {\Hc}. This reflects the smallness of $M'_a (G)$ and $M'_b (F)$  components for {\Ha},  as shown in Fig.~\ref{fig:SmallComponents}\,(a). 
Second, the torque for {\Ha} remains non-zero up to $H_{\bf a}^{\ast}$, whereas the torque for {\Hc} remains non-zero up to $H_{\bf c}^{\ast\ast}$. This again stems from the associated behaviors of $M'_a(G)$ and $M'_b(F)$ [see Figs.~\ref{fig:SmallComponents}\,(a) and \ref{fig:AnsatzResults}\,(f)]. 
Third, both torques show a non-monotonic behavior as a function of the field. The torque for {\Hc}, in particular, shows a characteristic sharp kink at $H_{\bf c}^\ast$, reflecting the first order transition between the low-field six-sublattice and the high-field two-sublattice state.
Importantly, this kink remains sharp even after we include the leading $1/S$ spin-wave corrections (blue line, see Appendix~\ref{app:SW}). A measurement of the torque can therefore give direct evidence for the transition at $H_{\bf c}^\ast$, and thus provide information for the value of $\Gamma$ via Eq.~(\ref{eq:Hacstar}).

Finally, for {\Hc}, the torque in the high-field phase scales as 
\be
H_{\bf c}^\ast \leq H\leq H_{\bf c}^{\ast\ast}:~~ \tau/H \propto \sqrt{1-\left(H/H_{\bf c}^{\ast\ast}\right)^2}\,.
\ee
Thus, a measurement of the torque at high fields can also be used to extract $H_{\bf c}^{\ast\ast}$ and, in turn, an independent constraint on the microscopic parameters $J$ and $\Gamma$ via Eq.~(\ref{eq:Hcstarstar}).

\vspace*{-0.3cm}
\section{Effect of thermal fluctuations \& classical $H$-$T$ phase diagram}\label{sec:FiniteT}	
\vspace*{-0.3cm}
To cross-check the above zero-temperature results from the classical ans\"atze and confirm the high-field thermal transition for {\Hc} mentioned above, we have performed classical Monte Carlo simulations using the standard Metropolis algorithm combined with the over-relaxation algorithm~\cite{Metropolis1953,Creutz1987}. The simulations were performed on finite-size clusters with a total number of sites ${\mc N}\!\in\!\{48,96,144,192,240,288\}$ and periodic boundary conditions. All considered systems, spanned by the unit vectors of the orthorhombic lattice, have at least three periods in the orthorhombic ${\bf a}$-direction in order to accommodate ${\bf Q}=2\hat{\bf a}/3$ order, see more details in Appendix~\ref{app:MC}. 
The results obtained by a thermal annealing down to $T\!=\!5$~K show that the total magnetization is almost indistinguishable from the predictions of the semi-analytical approach, lending strong support that the latter delivers quantitatively accurate results for the local physics of the problem. 

\begin{figure}[!t]
{\includegraphics[width=0.95\columnwidth]{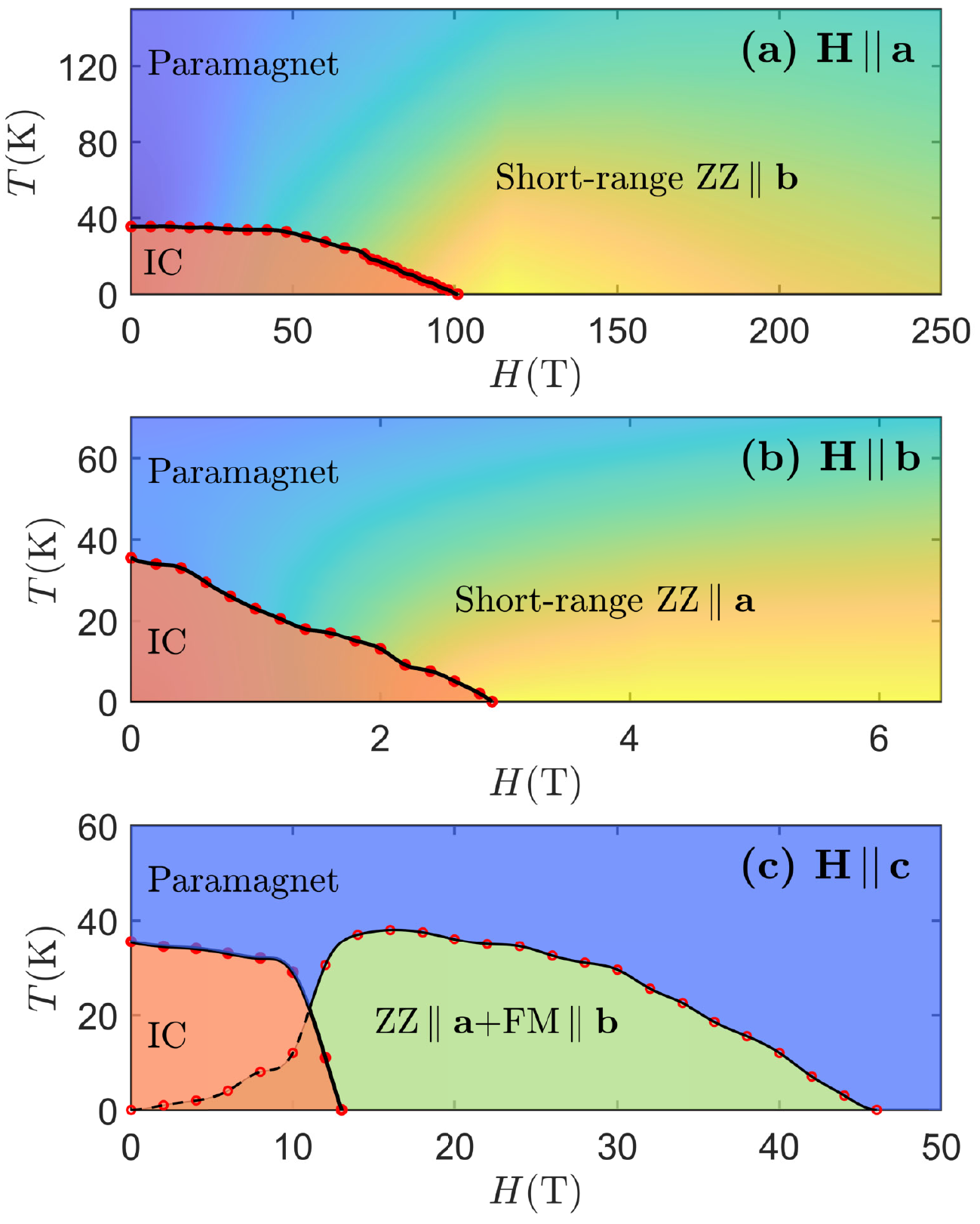}}
\caption{The field-temperature phase diagram obtained from MC simulations for field applied along (a) $\bf a$-, (b) $\bf b$-, and (c) $\bf c$-axis.} \label{fig:phaseD}
\end{figure}

Let us now turn to the classical $H$-$T$ phase diagrams, which are shown in Fig.~\ref{fig:phaseD} for the three orthorhombic directions. The boundary lines of the counter-rotating order (denoted by `IC') have been extracted by a finite-size analysis of the so-called Binder cumulant~\cite{Binder1993} (see Appendix~\ref{app:MC}),
\sbe
B_{\mc{O}_{{\bf Q}=2\hat{\bf a}/3}}= 1-\langle \mc{O}^4_{{\bf Q}=2\hat{\bf a}/3}\rangle \Big{/} \Big( 3\langle \mc{O}^2_{{\bf Q}=2\hat{\bf a}/3}\rangle^2 \Big)\,,
\see
of the equally-weighted combination of the three modulated static structure factor components (for all field directions):
\sbe
\mc{O}_{{\bf Q}=2\hat{\bf a}/3}=\sqrt{|M_{\bf a}(A)|^2+|M_{\bf b}(C)|^2+|M_{\bf c}(F)|^2}\,.
\see

For {\Hb}, the phase diagram contains two distinct phases, the high-$T$ paramagnetic phase and the low-$T$ counter-rotating order, which persists up to $H_{\bf b}^\ast\!\sim\!2.8$~T, see Fig.~\ref{fig:phaseD}\,(b). 

For {\Ha}, the counter-rotating order persists up to  very high fields ($H_{\bf a}^\ast\!\sim\!102$~T), see Fig.~\ref{fig:phaseD}\,(a), and is accompanied by the uniform orders $M'_a(G)$ and $M'_b(F)$ which are however extremely weak, see Fig.~\ref{fig:SmallComponents}\,(a). While these orders onset at the same field $H_{\bf a}^\ast$ as the modulated order at $T\!=\!0$, it is unclear whether this remains true for finite $T$. In fact, symmetry considerations alone tell us that the boundaries of the two types of orders can in general be different, as the they break different symmetries (the modulated order breaks translations whereas the uniform orders break $C_{2{\bf a}}$ and $\Theta C_{2{\bf b}}$, see Table~\ref{tab:symmetries}). Unfortunately, the smallness of $M'_a(G)$ and $M'_b(F)$ does not allow for an accurate numerical determination of their transition temperature line.

For {\Hc}, there are three distinct phases, see Fig.~\ref{fig:phaseD}\,(c). Apart from the paramagnetic and the modulated phase, there is a robust high-field order associated with $M'_a(G)$ and $M'_b(F)$, and the spontaneous breaking of $C_{2\bf c}$ and $\Theta C_{2\bf b}$ (see Table~\ref{tab:symmetries}). This phase coexists with the modulated order at low $H$ and $T$, but extends up to very high fields ($H_{\bf c}^{\ast\ast}\!\sim\!45$~T). Its boundary line has been extracted from the Binder cumulant $B_{\mc{O}_{{\bf Q}=0}}$ associated with
\sbe
\mc{O}_{{\bf Q}=0} = \sqrt{|M_{\bf a}'(G)|^2+|M_{\bf b}'(F)|^2}\,.
\see

Finally, the yellow shading in Figs.~\ref{fig:phaseD}\,(a) and (b) represents the variation of the magnitude of $|M_{\bf b}'(G)|$ and $|M_{\bf a}'(G)|$, respectively, from high values (intense yellow) at low $T$ to vanishing values (blue) at higher $T$. 

\vspace{-0.3cm}
\section{Discussion}\label{sec:Discussion}
\vspace{-0.3cm}
The study presented here provides a semi-analytical framework for the anisotropic response of $\beta$-$\text{Li}_2\text{IrO}_3$ under a magnetic field along the three orthorhombic directions. This framework is based on the minimal nearest-neighbor $J$-$K$-$\Gamma$ model~\cite{Lee2015,Lee2016,Ducatman2018,Rousochatzakis2018} and the hypothesis that the local correlations of the low-field incommensurate order can be captured by its closest commensurate approximant with the right symmetry~\cite{Ducatman2018}. The results are in qualitative agreement with almost all experimental facts collected so far, and we have shown how a quantitative agreement can also be reached by including quantum fluctuations. 

In addition, our analysis delivers a number of predictions which await experimental verification. 
First, the critical fields $H^\ast$ that mark the disappearance of the modulated order are highly anisotropic, in particular, $H_{\bf b}^\ast\!<\!H_{\bf c}^\ast\!\ll\!H_{\bf a}^\ast$. Such an anisotropic response, which is also evidenced in susceptibility~\cite{Ruiz2017,Majumder2019}, signifies a large separation of energy scales between $J$ and $\Gamma$. An explicit dependence of $H^\ast$ on these interactions is derived in this work [Eq.~(\ref{eq:Hacstar})] and can be used to extract the actual strength of $\Gamma$ (the value of $J$ is estimated $\sim 4$~K from the value of $H_{\bf b}^\ast$~\cite{Rousochatzakis2018}). Importantly, the dominant Kitaev coupling $K$ does not affect any of the critical fields,  partly because it is ferromagnetic. 

Second, for all orthorhombic directions our analysis reveals the presence of various intertwined uniform zigzag and  FM orders, some of which remain robust far above $H^\ast$. The physical origin of this robustness is related to the cross-coupling  terms of Eqs.~(\ref{eq:cross-coupling1}-\ref{eq:cross-coupling2}). Some of the uniform orders give rise to a finite torque signal and can thus be detected in a direct way. Alternatively, they can also be observed by magnetic X-ray diffraction~\cite{Ruiz2017}, or by local probes like NMR or $\mu$SR. 

Third, we have shown that the high-field response for {\Hc} is special, in that the disappearance of the modulated order at $H_{\bf c}^\ast$ restores only the translational symmetry and leaves some of the discrete symmetries broken. This implies the presence of a second thermal transition above $H_{\bf c}^\ast$, which is associated with the onset of the uniform orders $M'_a(G)$ and $M'_b(F)$. This transition can then be detected with thermodynamic measurements at high enough fields.

A natural extension of the present study is the investigation of the field-induced behavior of $\beta$-Li$_2$IrO$_3$ for general field directions, i.e., away from the orthorhombic axes. As it turns out, a semi-analytical description can be also obtained for fields in the $ab$- and $bc$- planes~\cite{Mengqun2}. The emerging picture reveals a remarkable interplay of the various modulated and uniform orders and rich anisotropic phase diagrams, the details of which will be given elsewhere~\cite{Mengqun2}. We can, however, comment on one particular aspect related to the torque signal discussed in Sec.~\ref{sec:Torque}.  As mentioned there, the predicted torque signals are proportional to the quantity $\xi = g_{bb} M'_b(F)+g_{ab}M'_a(G)$, whose sign is chosen spontaneously by the system for {\Ha} or {\Hc}. However, adding an infinitesimal field along ${\bf b}$ will actually fix the sign of $\xi$, since the two are directly coupled to each other. This simple argument shows that, as a function of the angle in the $ab$- or $bc$-planes, the torque will show an abrupt reversal when the field passes through the ${\bf a}$ and ${\bf c}$ axes, respectively. Such a first-order transition scenario could also be relevant for the explanation of the sawtooth-like torque anomalies observed experimentally in the closely related compound $\gamma$-Li$_2$IrO$_3$~\cite{Modic2017,Modic2018} (see also \cite{Winter2019}).

Finally, we would like to touch upon an aspect that may be relevant for the interpretation of the phase transition reported recently around 100 K~\cite{Ruiz2019}.  As discussed in Ref.~\cite{Ducatman2018}, the zero-field and zero-temperature configuration contains the uniform orders $M'_a(G)$ and $M'_b(F)$, in addition to the modulated order. Given that the two types of order break different symmetries (the modulated order breaks translations whereas the uniform orders break $C_{2{\bf a}}$ and $C_{2{\bf c}}$~\cite{Ducatman2018}) one generally expects that the two types of order onset at different temperatures. In particular, we have checked numerically (unpublished) that the modulated period-3 six-sublattice order carries a pseudo-Goldstone  low-energy mode, similar to other incommensurate phases in related models~\cite{Mengqun2019,Choi_2013,Choi2013}. On the other hand, the energy barrier associated with flipping the signs of the uniform orders $M'_a(G)$ and $M'_b(F)$ gives rise to a finite energy gap. It is then plausible that the uniform  orders onset at a higher temperature $T_{\text{uni}}$ compared to $T_N$. While the smallness of the uniform orders does not allow to check this numerically with Monte Carlo, the cross-coupling term of Eq.~(\ref{eq:cross-coupling1}) suggests that $T_{\text{uni}}$ could scale with $\Gamma$. In such a scenario, a field along the ${\bf b}$ axis will turn the zero-field line extending from $T\!=\!0$ up to $T\!=\!T_{\text{uni}}$ into a line of first order transitions, because the field couples directly to $M'_b(F)$ (and to $M'_a(G)$ via $g_{ab}$). For very low fields, the proximity to this first-order line would then give rise to hysteresis effects, similar to those observed in Ref.~ \cite{Ruiz2019}. The actual details of this scenario (in particular, the connection of the measured torque signals with the ones we report here at zero-temperature), remain to be explored.

\vspace*{0.3cm} 
\noindent{\it  Acknowledgments:} 
We thank J. Analytis, J. Betouras, A. Ruiz and A. A. Tsirlin  for helpful discussions. This work was supported by the U.S. Department of Energy, Office of Science, Basic Energy Sciences under Award No. DE-SC0018056. We also acknowledge the support of the Minnesota Supercomputing Institute (MSI) at the University of Minnesota. N.B.P. also acknowledges  the hospitality of the Aspen Center for Physics supported by National Science Foundation grant PHY-1607611, where part of the work on this manuscript has been done.



\appendix
\titleformat{\section}{\normalfont\bfseries\filcenter}{Appendix~\thesection:}{0.25em}{}

\vspace*{-0.3cm}
\section{Static structure factors}\label{app:SSF}
\subsection{Definitions and conventions}\label{app:SSFconventions}
\vspace*{-0.3cm}
Each orthorhombic unit cell contains four primitive cells (labeled by $i=1$-$4$), and each primitive cell contains four spin sites (labeled by $\nu=1$-$4$). Each site can then be labeled by the position ${\bf R}$ of the orthorhombic unit cell, the position $\bs{\rho}_i$ of the primitive unit cell (relative to ${\bf R}$), and the position ${\bf p}_\nu$ of the spin sublattice (relative to $\bs{\rho}_i$). The physical position of each site can then be written as 
\be
{\bf r}_{{\bf R},i,\nu} = {\bf R}+\bs{\rho}_i+{\bf p}_\nu\,.
\ee
The Fourier transform of the $\nu$-th spin sublattice is defined as 
\sbe\label{eq:SofQ}
{\bf S}_{\nu}({\bf Q}) =\frac{1}{\mc{N}/16} \sum_{{\bf R}, i} e^{i {\bf Q}\cdot({\bf R}_i+\bs{\rho}_i+{\bf p}_\nu)} {\bf S}_{{\bf R}, i, \nu}\,,
\see
where $\mc{N}$ is the total number of spins and ${\bf Q}$ belongs to the reciprocal space of the orthorhombic Bravais lattice. 

The modulated ${\bf Q}\!=\!2\hat{\bf a}/3$ components of the static structure factor are defined as [see Eq.~(\ref{eq:symmetrybasis})] 
\sbe\label{eq:Ms}
\!\!\!\begin{pmatrix}
i\,{\bf M}(A)\\
i\,{\bf M}(C)\\
{\bf M}(F)\\
i\,{\bf M}(G)\\
\end{pmatrix} \!\equiv\! \frac{1}{4}
\begin{pmatrix}
{\bf S}_1({\bf Q})-{\bf S}_2({\bf Q})-{\bf S}_3({\bf Q})+{\bf S}_4({\bf Q})\\
{\bf S}_1({\bf Q})+{\bf S}_2({\bf Q})-{\bf S}_3({\bf Q})-{\bf S}_4({\bf Q})\\
{\bf S}_1({\bf Q})+{\bf S}_2({\bf Q})+{\bf S}_3({\bf Q})+{\bf S}_4({\bf Q})\\
{\bf S}_1({\bf Q})-{\bf S}_2({\bf Q})+{\bf S}_3({\bf Q})-{\bf S}_4({\bf Q})
\end{pmatrix}_{{\bf Q}=2\hat{\bf a}/3}\!\!\!\!.
\see
Note that the extra prefactors of $i$ in the definitions of ${\bf M}(A)$, ${\bf M}(C)$ and ${\bf M}(G)$ have been inserted to follow the convention of Ref.~[\onlinecite{Biffin2014a}], while the normalization prefactor $1/4$ in the right hand side of Eq.~(\ref{eq:Ms}) sets the maximum possible magnitude of the various components to $S$.
Similarly, the uniform ${\bf Q}\!=\!0$ components of the static structure factor are defined as
\sbe\label{eq:Mps}
\begin{pmatrix}
{\bf M}'(A)\\
{\bf M}'(C)\\
{\bf M}'(F)\\
{\bf M}'(G)\\
\end{pmatrix} \equiv \frac{1}{4}
\begin{pmatrix}
{\bf S}_1({\bf 0})-{\bf S}_2({\bf 0})-{\bf S}_3({\bf 0})+{\bf S}_4({\bf 0})\\
{\bf S}_1({\bf 0})+{\bf S}_2({\bf 0})-{\bf S}_3({\bf 0})-{\bf S}_4({\bf 0})\\
{\bf S}_1({\bf 0})+{\bf S}_2({\bf 0})+{\bf S}_3({\bf 0})+{\bf S}_4({\bf 0})\\
{\bf S}_1({\bf 0})-{\bf S}_2({\bf 0})+{\bf S}_3({\bf 0})-{\bf S}_4({\bf 0})
\end{pmatrix} \,.
\see

\vspace*{-0.3cm}
\subsection{Local spin length constraints in terms of structure factors}\label{app:Constraints}
\vspace*{-0.3cm}
We now show that the intensity sum rule [Eq.~(\ref{eq:ISR0})] is a direct consequence of the local spin length constraints. Inverting Eq.~(\ref{eq:SofQ}) we get
\sbe
\begin{array}{l}
{\bf S}_{{\bf R},\nu,i}  =  \sum_{{\bf Q}\in\text{BZ}} e^{-i {\bf Q}\cdot({\bf R}+\bs{\rho}_i+{\bf p}_{\nu})} {\bf S}_{\nu}({\bf Q})\,,
\end{array}
\see
where the sum over ${\bf Q}$ is over the first Brillouin zone of the orthorhombic Bravais lattice. The local spin length constraints then take the form
\sbea
\begin{array}{l}
{\bf S}_{{\bf R},\nu,i} ^2\!=\! 
\sum_{{\bf (q-Q)}\in\text{BZ}}
e^{-i {\bf q}\cdot({\bf R}+\bs{\rho}_i+{\bf p}_{\nu})}
\sum_{{\bf Q}\in\text{BZ}} {\bf S}_{\nu}({\bf q}-{\bf Q})\cdot
{\bf S}_{\nu}({\bf Q})\!=\!S^2,
\end{array}
\seea
which holds for all ${\bf R}$, $\nu$ and $i$ if we require
\sbea
\begin{array}{l}
f_{\nu}({\bf q})\equiv\sum_{{\bf Q}\in\text{BZ}}  {\bf S}_{\nu}({\bf q}-{\bf Q})\cdot
{\bf S}_{\nu}({\bf Q})=S^2 \delta_{{\bf q},0}\,, \forall \nu\!=\!1\!-\!4\,.
\end{array}
\seea
The intensity sum rule derives from the ${\bf q}\!=\!0$ part, namely 
\sbea\label{eq:sumSofQsq}
\begin{array}{l}
\sum_{{\bf Q}\in\text{BZ}}  |{\bf S}_{\nu}({\bf Q})|^2=S^2\,, \forall \nu\!=\!1\!-\!4\,.
\end{array}
\seea
To see this let us take the general form of Eqs.~(\ref{eq:Ms}) and (\ref{eq:Mps}) for any ${\bf Q}$, 
\sbe\label{eq:MofQgeneral}
\begin{pmatrix}
f {\bf M}_{\bf Q}(A)\\
f {\bf M}_{\bf Q}(C)\\
{\bf M}_{\bf Q}(F)\\
f {\bf M}_{\bf Q}(G)\\
\end{pmatrix} \equiv \frac{1}{4}
\begin{pmatrix}
{\bf S}_1({\bf Q})-{\bf S}_2({\bf Q})-{\bf S}_3({\bf Q})+{\bf S}_4({\bf Q})\\
{\bf S}_1({\bf Q})+{\bf S}_2({\bf Q})-{\bf S}_3({\bf Q})-{\bf S}_4({\bf Q})\\
{\bf S}_1({\bf Q})+{\bf S}_2({\bf Q})+{\bf S}_3({\bf Q})+{\bf S}_4({\bf Q})\\
{\bf S}_1({\bf Q})-{\bf S}_2({\bf Q})+{\bf S}_3({\bf Q})-{\bf S}_4({\bf Q})
\end{pmatrix} \,,
\see
where $f\!=\!i$ or $1$ [see Eqs.~(\ref{eq:Ms}) and (\ref{eq:Mps})]. Squaring each row and adding them up gives
\sbea
\!\!|{\bf M}_{\bf Q}(A)|^2\!+\!|{\bf M}_{\bf Q}(C)|^2\!+\!|{\bf M}_{\bf Q}(F)|^2\!+\!|{\bf M}_{\bf Q}(G)|^2 \!=\! \frac{1}{4} \!\!\sum_{\nu=1-4}\!\!|{\bf S}_\nu({\bf Q})|^2,~~~
\seea
which in conjunction with Eq.~(\ref{eq:sumSofQsq}) gives
\sbe
\!\!\sum_{{\bf Q}\in\text{BZ}}\!\left(|{\bf M}_{\bf Q}(A)|^2\!+\!|{\bf M}_{\bf Q}(C)|^2\!+\!|{\bf M}_{\bf Q}(F)|^2\!+\!|{\bf M}_{\bf Q}(G)|^2\right) \!=\! S^2\,.
\see
The only ${\bf Q}$ vectors inside the first Brillouin zone of the orthorhombic lattice that contribute to this sum are the ones corresponding to ${\bf Q}\!=\!\pm \frac{2}{3}\hat{\bf a}$ and ${\bf Q}\!=\!0$, which leads to the intensity sum rule Eq.~(\ref{eq:ISR0}).

Note that the above analysis can be carried out for quantum spins as well, in which case the various spin-spin correlations, such as ${\bf S}_i\cdot{\bf S}_j$, must be replaced with the corresponding expectation values $\langle{\bf S}_i\cdot{\bf S}_j\rangle$ in the quantum-mechanical ground state of the system, and ${\bf S}_i\cdot {\bf S}_i$ becomes $S(S+1)$.

According to the above, the Bragg peak intensity sum rule is very general and does not depend on the particular values of the microscopic parameters. This generality was missed in \cite{Rousochatzakis2018}, because the components $M_a$, $M_a'$, $M_c$ and $M_c'$ defined there differ by a relative prefactor of $\sqrt{2}$ from the ones defined here, while this is not the same for the components $M_b$ and $M_b'$. As a result, the quantity $I_{\text{tot}}$ defined in \cite{Rousochatzakis2018} does not correspond to the intensity defined here, which is why that quantity satisfies the sum rule only for sufficiently small $J$ (compare in particular the two panels of Fig.~4 of \cite{Rousochatzakis2018}).

\vspace*{-0.3cm}
\section{Auxiliary information for the various ans\"atze}\label{app:ansatze}

\vspace*{-0.3cm}
\subsection{Field along the crystallographic $\bf b$-axis.}\label{app:Halongb}
\vspace*{-0.3cm}
\subsubsection{Low-field phase for {\Hb}}
\vspace*{-0.3cm}
According to Table~\ref{tab:ansatze}, the low-field ansatz for {\Hb} reads 
\sbe\label{Kstate}
\begin{array}{ll}
\vec{A}=S[x_1,y_1,z_1],~~	& \vec{A'}=S[y_1,x_1,z_1], \\[1ex]
\vec{B}=S[-y_1,-x_1,z_1],~~ & \vec{B'}=S[-x_1,-y_1,z_1], \\[1ex]
\vec{C}=S[-x_2,x_2,-z_2],~~ & \vec{C'}=S[x_2,-x_2,-z_2],
\end{array}
\see
where $x_1$, $x_2$, $y_1$, $z_1$ and $z_2$ denote Cartesian components of spins. Due to the spin-length constraints $x_1^2\!+\!y_1^2\!+\!z_1^2\!=\!1$ and $2x_2^2\!+\!z_2^2\!=\!1$, only three out of these five parameters are independent. 
The state can also be parametrized in terms of the five symmetry-resolved static structure factor components $M_a(A)$, $M_b{C}$, $M_c(F)$, $M'_a(G)$ and $M'_b(F)$, which are related to the Cartesian components by 
\sbe\label{eq:SFvsCartesianb}
\begin{array}{l}
M_a(A)=iS(x_1+2x_2-y_1)/(3\sqrt{2}) \,, \\
M_b(C)=iS(z_1+z_2)/3 \,,  \\
M_c(F)=iS(x_1+y_1)/\sqrt{6} \,,  \\
M_a'(G)=-2S(x_1-y_1-x_2)/(3\sqrt{2}) \,, \\
M_b'(F)=-S(2z_1-z_2)/3 \,. 
\end{array}
\see

Out of the five structure factor components only three are independent, as there are two spin length constraints. One of them is the Bragg peak intensity sum rule, 
\sbe
\!\!I_{\text{tot}}\!=\!2\big[|M_a(A)|^2\!+\!|M_b(C)|^2\!+\!|M_c(F)|^2\big] \!+\! |M'_a(G)|^2\!+\!|M'_b(F)|^2 \!=\! S^2.
\see
The second constraint reads 
\sbe\label{eq:CrossCouplingTerms}
\!\!\!\!|M_c(F)|^2\!=\!|M_a(A)|^2\!+\!|M_b(C)|^2\!-\!2i\big[M_a(A)M'_a(G)\!+\!M_b(C)M'_b(F)\big]\,.
\see 
This illustrates how the local spin length constraints can lead to effective cross-coupling terms between the modulated and uniform components, i.e., the terms $M_a(A)M'_a(G)\!+\!M_b(C)M'_b(F)$.

The total energy per site is given by
\sbe\label{eq:energy_b1}
\renewcommand{\arraystretch}{1.2}
\begin{array}{l}
E_{\bf b}/\mc{N} \!=\!\sfrac{1}{6} S^2\Big\{K [3\!-\!2(y_1\!-\!x_2)^2 ]
\!+\! 2 \Gamma [1\!-\!z_1^2 \!+\! x_2^2\!+\! 2 (y_1z_1\\
~~+x_2z_1\!+\! x_1 z_2) ] 
\!+\! J [1\!+\!2 (z_1-z_2)^2\!-\! 4 x_1 x_2 \!+\! 4 (x_1\!+\!x_2) y_1 ] \Big\} \\
~~ - \sfrac{1}{3}S\mu_B H [\sqrt{2} g_{ab} (x_1 - x_2 - y_1) + g_{bb} (-2 z_1 + z_2)]\,.
\end{array}
\see
where $\mc{N}$ is the total number of spin sites. In terms of the structure factor components, $E_{\bf b}$ takes the form 
\sbea\label{eq:energy_b2}
\renewcommand{\arraystretch}{1.2}
\begin{array}{l}
E_{\bf b}/\mc{N} \!=\!
\eta_{aA} M_a(A)^2 \!+\!\eta_{bC} M_b(C)^2 \!+\!\eta_{cF} M_c(F)^2 \!+\! \eta'_{aG} M'_a(G)^2
\!+\!\eta'_{bF} M'_b(F)^2\\
~- \sqrt{2} \Gamma \big[ M_a(A) M_b(C) \!+\! \sqrt{3} M_b(C) M_c(F)
\!+\! M'_a(G) M'_b(F)\big] \\
~- \sqrt{3}K M_a(A) M_c(F) 
\!-\!\mu_B H  \big[ g_{bb} M'_b(F)-g_{ab} M'_a(G) \big]\,,
\end{array}
\seea
where
\sbea
\renewcommand{\arraystretch}{1.2}
\begin{array}{c}
\eta_{aA}=-\Gamma\!\!+2J\!+\!K/2,~~
\eta_{bC}=-K,~~
\eta_{cF}= -(\Gamma\!+\!2J\!+\!K/2),\\
\eta'_{aG}=\sfrac{1}{2}(\Gamma\!+\!J\!+\!K),~~
\eta'_{bF}=\sfrac{1}{2}(3J\!+\!K)\,.
\end{array}
\seea
Note that while there are no cross-coupling terms between the modulated and the uniform components, such terms arise from the spin-length constraints, as shown in Eq.~(\ref{eq:CrossCouplingTerms}).

\vspace*{-0.3cm}
\subsubsection{High-field phase for {\Hb}}
\vspace*{-0.3cm}
For $H\!\ge\!H_{\bf b}^*$ the Cartesian components satisfy the relations 
\sbe
x_1\!=\!-y_1\!=\!-x_2\,,~~~ z_2\!=\!-z_1\,, 
\see
and we are left with the two-sublattice ansatz [see Table~\ref{tab:ansatze}] 
\sbe
\begin{array}{l}
{\bf A}={\bf B}={\bf C}\equiv {\bf F} =S [x_1,-x_1,z_1]\,,\\
{\bf A}'={\bf B}'={\bf C}'\equiv {\bf F}' = S [-x_1,x_1,z_1]\,,
\end{array}
\see
with $z_1^2\!=\!1\!-\!2x_1^2$, $x_1\!>\!0$ and $z_1\!<\!0$. 
In this phase, the modulated components $M_a(A)$, $M_b(C)$ and $M_c(F)$ vanish identically, and we are left with the two uniform components     
\sbe
M_a'(G)=-\sqrt{2} S x_1,~~ M_b'(F)=-S z_1\,,
\see
subject to the constraint $M_b'(F)^2\!+\!M_a'(G)^2\!=\!S^2$. 
The total energy Eq.~(\ref{eq:energy_b2}) becomes
\sbe\label{eq:energy_b3}
\renewcommand{\arraystretch}{1.2}
\begin{array}{l}
E_{\bf b}/\mc{N} \!=\! \eta'_{bF} M_b'(F)^2 \!+\! \eta'_{aG} M_a'(G)^2 
-\sqrt{2} \Gamma ~M_a'(G)~ M_b'(F)\\
~~~~ 
- \mu_B H \big[ g_{bb} M_b'(F) - g_{ab} M_a'(G) \big] \,.
\end{array}
\see
Minimizing gives the following relation between the magnitude of the field $H$, and the components $x_1$ and $z_1$:
\sbe
\frac{\mu_B H}{2S} \!=\! \frac{\Gamma(4x_1^2\!-\!1)\!-\!(2J\!-\!\Gamma) x_1 z_1}{2g_{bb}x_1+\sqrt{2}g_{ab} z_1},\qquad H\!\ge\!H_{\bf b}^\ast~.
\see
In the limit of very large field, $H\to\infty$, 
\sbe
\sqrt{2}x_1\!\to\!\frac{g_{ab}}{\sqrt{g_{bb}^2\!+\!g_{ab}^2}}\,, ~~
z_1\!\to\!-\frac{g_{bb}}{\sqrt{g_{bb}^2\!+\!g_{ab}^2}}\,.
\see
 
Note that the cross-coupling term $-\sqrt{2}\Gamma M_a'(G) M_b'(F)$ in Eq.~(\ref{eq:energy_b3}) favors opposite signs of $M_a'(G)$ and $M_b'(F)$, given that $\Gamma\!<\!0$. And since the magnetic field favors a positive $M_b'(F)$, it follows that $\Gamma$ favors a negative $M_a'(G)$ (the term $\propto g_{ab}$ in Eq.~(\ref{eq:energy_b3}) also favors a negative $M_a'(G)$ if $g_{ab}\!>\!0$; otherwise $M_a'(G)$ must turn positive at high enough fields). In other words, the sign of the zigzag component along ${\bf a}$ is fixed by the field.

\vspace*{-0.3cm}
\subsection{Field along the crystallographic $\bf a$-axis.}\label{app:Halonga}
\vspace*{-0.3cm}
\subsubsection{Low-field phase for {\Ha}}
\vspace*{-0.3cm}
According to Table~\ref{tab:ansatze}, the low-field ansatz for {\Ha} reads 	
\sbe
\begin{array}{ll}
{\bf A}=S[x_1,y_1,z_1]\,,~~	& \vec{A'}=S[y_2,x_2,z_2]\,, \\[1ex]
{\bf B}=S[-y_1,-x_1,z_1]\,,~~ &\vec{B'}=S[-x_2,-y_2,z_2]\,, \\[1ex]
{\bf C}=S[-x_3,x_3,-z_3]\,,~~ &\vec{C'}=S[x_4,-x_4,-z_4]\,.
\end{array}
\see
Here we have ten Cartesian components which obey the four constraints $x_1^2\!+\!y_1^2\!+\!z_1^2\!=\!1$, $x_2^2\!+\!y_2^2\!+\!z_2^2\!=\!1$, $2x_3^2\!+\!z_3^2\!=\!1$, and $2x_4^2\!+\!z_4^2\!=\!1$. Therefore, only six Cartesian components are independent. 
Note that for $H\!=\!0$, the minimum satisfies the relations $x_1\!=\!x_2$, $y_1\!=\!y_2$, $z_1\!=\!z_2$, $x_3\!=\!x_4$, and $z_3\!=\!z_4$, and the ansatz reduces to the form  given in Eq.~(\ref{Kstate}).

The state can also be described in terms of ten structure factor components. Among these, the first five are the ones we encounter at zero field (and for finite fields along ${\bf b}$). The remaining five include three field-induced modulated components $M_a(C)$, $M_b(A)$ and $M_c(G)$, and two field-induced uniform components $M'_a(F)$ and $M'_b(G)$. Their dependence on the Cartesian components is
\sbe\label{eq:SFvsCartesiana}
\renewcommand{\arraystretch}{1.2}
\begin{array}{l}
M_a(A)=i S(x_1+x_2-y_1-y_2+2x_3+2x_4)/(6\sqrt{2}) \,, \\
M_b(C)=i S(z_1+z_2+z_3+z_4)/6 \,, \\
M_c(F)=i S(x_1+y_1+x_2+y_2)/(2\sqrt{6}) \,,\\
M_a'(G)=-S(x_1+x_2-x_3-x_4-y_1-y_2)/(3\sqrt{2}) \,, \\
M_b'(F)=-S(2z_1+2z_2-z_3-z_4)/6 \,,\\
\hline
M_a(C)=-i S (x_1-x_2-y_1+y_2+2x_3-2x_4)/(6\sqrt{2}) \,,\\
M_b(A)=-i S(z_1-z_2+z_3-z_4)/6 \,, \\
M_c(G)=-i S(x_1+y_1-x_2-y_2)/(2\sqrt{6}) \,, \\
M_a'(F)=-S(x_2-x_1+x_3-x_4+y_1-y_2)/(3\sqrt{2}) \,,\\
M_b'(G)=-S(2z_2-2z_1+z_3-z_4)/6 \,.
\end{array}
\see
The ten structure factor components obey four constraints. One of them is the Bragg peak intensity sum rule given in Eq.~(\ref{eq:ISRa}).
The remaining three constraints involve various types of effective cross-coupling terms, similar to the ones we have seen in Eq.~(\ref{eq:CrossCouplingTerms}). For example, one of these constraints reads: 
\sbe
\renewcommand{\arraystretch}{1.2}
\!\!\!\!\!\!\!\!\!\begin{array}{c}
2 \big[ M_a(A) M_a(C)+M_b(A) M_b(C)+ i M_c(F) M_c(G)\Big]\\
~~=M'_a(F) M'_a(G)+M'_b(F) M'_b(G)
\end{array}
\see

The  total energy of the system reads
\sbe\label{eq:energy_a1}
\renewcommand{\arraystretch}{1.2}
\!\!\!\!\!\!\begin{array}{l}
E_{\bf a}/\mc{N}\!=\!\sfrac{1}{6}S^2\Big\{K [x_1^2\!+\!x_2^2\!+\!2(x_3y_1\!+\!x_4y_2\!+\!z_1z_2)\!+\!z_3z_4]
\!+\!2 \Gamma [x_1x_2\\
~+x_3x_4\!+\!y_1y_2\!+\!x_1z_3\!+\!x_3z_1\!+\!x_2z_4\!+\!x_4z_2\!+\!y_1z_1\!+\!y_2z_2] \\ 
~+2J [1\!-\!x_1x_3\!-\!x_2x_4\!-\!x_3x_4\!+\!x_1y_2\!+\!x_2y_1\!+\!x_3y_1\!+\!x_4y_2\\
~+z_1z_2\!-\!z_1z_3\!-\!z_2z_4\!+\!z_3z_4] \Big\}
\!-\!\sfrac{1}{6} S\mu_B H \big\{g_{ab} \big[2(z_2\!-\!z_1)\\
~+z_3\!-\!z_4\big]
\!+\!\sqrt{2}g_{aa} (x_1\!-\!x_2\!-\!x_3\!+\!x_4\!-\!y_1\!+\!y_2)\big\} \,,
\end{array}
\see
or, in terms of the static structure factor components, 
\sbea\label{eq:energy_a2}
\renewcommand{\arraystretch}{1.2}
\!\!\!\!\!\!\!\!\begin{array}{l}
E_{\bf a}/\mc{N} \!=\! 
\eta_{aA} M_a(A)^2 \!+\! \eta_{bC} M_b(C)^2 \!+\! \eta_{cF} M_c(F)^2
\!+\! \eta_{aC} M_a(C)^2 \\
~+\eta_{bA} M_b(A)^2
\!+\! \eta_{cG} M_c(G)^2 \!+\! \eta'_{aG} M'_a(G)^2 \!+\! \eta'_{bF} M'_b(F)^2\!+\! \eta'_{aF} M'_a(F)^2 \\
~+\eta'_{bG} M'_b(G)^2
\!-\! \sqrt{2} \Gamma \big[ 
M_a(C) M_b(A)\!+\!M_a(A) M_b(C) \\
~ +\sqrt{3} M_b(C) M_c(F)\!+\!\sqrt{3}  i M_b(A) M_c(G)  \!+\!M'_a(G) M'_b(F) \\
~+ M'_a(F) M'_b(G) \big]
\!-\! \sqrt{3} K \big[ M_a(A) M_c(F) \!+\! i M_a(C) M_c(G) \big]\\
~ -\mu_B H \big[ g_{aa} M'_a(F) \!-\! g_{ab} M'_b(G)\big]\,,
\end{array}
\seea
where we have introduced 
\sbea
\renewcommand{\arraystretch}{1.2}
\begin{array}{c}
\eta_{aC}=\Gamma+K/2,~~
\eta_{bA}=2J+K,~~
\eta_{cG}= -\Gamma+K/2,\\
\eta'_{aF}=\sfrac{1}{2}(-\Gamma + 3 J + K),~~
\eta'_{bG}=\sfrac{1}{2}(J-K) \,.
\end{array}
\seea

\vspace*{-0.3cm}
\subsubsection{High-field phase for {\Ha}}
\vspace*{-0.3cm}
For $H\!\geq\!H_{\bf a}^\ast$ the Cartesian components satisfy the relations 
\sbe
\renewcommand{\arraystretch}{1.2}
\begin{array}{c}
x_1\!=\!-x_2\!=\!-x_3\!=\!x_4\!=\!-y_1\!=\!y_2\,,\\[1ex]
z_1\!=\!-z_2\!=\!-z_3\!=\!z_4\,,
\end{array}
\see
and we are left with the two-sublattice ansatz  [see Table~\ref{tab:ansatze}] 
\sbe
\renewcommand{\arraystretch}{1.2}
\begin{array}{l}
{\bf A}={\bf B}={\bf C}\equiv {\bf F} =S [x_1,-x_1,z_1]\,,\\
{\bf A}'={\bf B}'={\bf C}'\equiv {\bf F}' = S [x_1,-x_1,-z_1]\,,
\end{array}
\see
with $2x_1^2\!+\!z_1^2\!=\!1$, $x_1\!>\!0$  and $z_1\!<\!0$. 
The only static structure factor components surviving for $H\!\geq\!H_{\bf a}^\ast$ are the uniform components $M'_b(G)$ and $M'_a(F)$,
\sbe
M_a'(F) = \sqrt{2}S x_1\,,~~~ M_b'(G) = S z_1\,,
\see
subject to the constraint $M_a'(F)^2\!+\!M_b'(G)^2\!=\!S^2$, and the total energy Eq.~(\ref{eq:energy_a2}) becomes
\sbea\label{eq:energy_a3}
\renewcommand{\arraystretch}{1.2}
\begin{array}{l}
E_{\bf a} / N \!=\! \eta'_{aF} M_a'(F)^2\!+\!\eta'_{bG} M_b'(G)^2\!-\!\sqrt{2}\Gamma~M_a'(F)~M_b'(G) \\
~~~~
-\mu_B H \left( g_{aa} M_a'(F)-g_{ab} M_b'(G)\right)\,.
\end{array}
\seea
Minimizing the total energy for $H\!\ge\!H_{\bf a}^\ast$ gives the following relation between $H$, $x_1$ and $z_1\!=\!-\sqrt{1-2x_1^2}$:
\sbe
\frac{\mu_B H}{2S}\!=\!\frac{\Gamma(4x_1^2\!-\!1)\!-\![\Gamma\!-\!2(J\!+\!K)] x_1 z_1}{2g_{ab}x_1\!+\!\sqrt{2}g_{aa} z_1}\,.
\see
In the limit of  $H\to \infty$, we get
\sbe
\sqrt{2} x_1\!\to\!\frac{g_{aa}}{\sqrt{g_{aa}^2\!+\!g_{ab}^2}}\,,~~~
z_1\!\to\!-\frac{g_{ab}}{\sqrt{g_{aa}^2\!+\!g_{ab}^2}}\,.
\see

Note that the cross-coupling term $-\sqrt{2}\Gamma M_b'(G) M_a'(F)$ in Eq.~(\ref{eq:energy_a3}) favors opposite signs of $M_b'(G)$ and $M_a'(F)$, since $\Gamma\!<\!0$. And given that the magnetic field favors a positive $M_a'(F)$, it follows that $M_b'(G)$ is negative (consistent with the term $\propto g_{ab}$ if $g_{ab}\!>\!0$). Therefore the sign of the zigzag component along ${\bf b}$ is fixed by the field.

\vspace*{-0.3cm}
\subsection{Field along the crystallographic $\bf c$-axis.}\label{app:Halongc}
\vspace*{-0.3cm}
\subsubsection{Low-field phase for {\Hc}}
\vspace*{-0.3cm}
According to Table~\ref{tab:ansatze}, the low-field ansatz for {\Hc} reads 
\sbe
\renewcommand{\arraystretch}{1.2}
\begin{array}{ll}
{\bf A}=S[x_1,y_1,z_1],~~	&\vec{A'}=S[y_1,x_1,z_1], \\
{\bf B}=S[-y_2,-x_2,z_2],~~ &\vec{B'}=S[-x_2,-y_2,z_2],\\
{\bf C}=S[-y_3,x_3,-z_3],~~ &\vec{C'}=S[x_3,-y_3,-z_3]\,.
\end{array}
\see
The nine Cartesian components obey three spin-length constraints, $x_1^2\!+\!y_1^2\!+\!z_1^2\!=\!1$, $x_2^2\!+\!y_2^2\!+\!z_2^2\!=\!1$, and $x_3^2\!+\!y_3^2\!+\!z_3^2\!=\!1$, and therefore only six components are independent. 
Note that at zero field, the minimum satisfies the relations $x_1\!=\!x_2$, $y_1\!=\!y_2$, $z_1\!=\!z_2$, $x_3\!=\!y_3$, and the ansatz Eq.~(\ref{Kstate}) is again restored. 

The state can also be parametrized by nine static structure factor components. Five of them are the ones we encounter at zero field (or for fields along ${\bf b}$). The remaining four include three field-induced modulated components $M_a(G)$, $M_b(F)$ and $M_c(C)$, and the uniform field-induced component $M'_c(F)$. The dependence on the Cartesian components is
\sbe\label{eq:SFvsCartesianc}
\renewcommand{\arraystretch}{1.2}
\begin{array}{l}
M_a(A)=i S(x_1+x_2+2x_3-y_1-y_2+2y_3)/(6\sqrt{2}) \,,\\
M_b(C)=i S(z_1+z_2+2z_3)/6 \,,\\
M_c(F)=i S(x_1+x_2+y_1+y_2)/(2\sqrt{6}) \,,\\
M_a'(G)=-S(x_1+x_2-x_3-y_1-y_2-y_3)/(3\sqrt{2}) \,,\\
M_b'(F)=-S(z_1+z_2-z_3)/3 \,,\\
\hline
M_a(G)=- S (x_1-x_2-y_1+y_2)/(2\sqrt{6}) \,,\\
M_b(F)= -i S(z_1-z_2)/(2 \sqrt{3}) \,, \\
M_c(C)= -i S(x_1-x_2+y_1-y_2-2x_3+2y_3)/(6\sqrt{2}) \,,\\
M_c'(F)=-S(x_1-x_2+x_3+y_1-y_2-y_3)/(3\sqrt{2}) \,.\\
\end{array}
\see
The nine static structure factor components obey three constraints. One of them is the Bragg peak intensity sum rule, which here reads
\sbe\label{eq:ISRc}
\renewcommand{\arraystretch}{1.2}
\!\!\!\!\begin{array}{c}
\!\!I_{\text{tot}}\!=\!2\Big\{|M_a(A)|^2\!+\!|M_b(C)|^2\!+\!|M_c(F)|^2+
|M_a(G)|^2\!+\!|M_b(F)|^2\\
~+|M_c(C)|^2\Big\}
+ |M'_a(G)|^2\!+\!|M'_b(F)|^2+ |M'_c(F)|^2
 \!=\! S^2.
\end{array}
\see
The total energy is given by
\sbe\label{eq:energy_c1}
\renewcommand{\arraystretch}{1.2}
\begin{array}{l}
E_{\bf c}/\mc{N}\!=\!\sfrac{1}{6}S^2\Big\{K [x_1^2\!+\!x_2^2\!+\!z_1^2\!+\!z_2^2\!+\!z_3^2\!+\!2x_3y_1\!+\!2y_2y_3] \\
~+\Gamma [x_1^2\!+\!x_2^2\!+\!x_3^2\!+\!y_1^2\!+\!y_2^2\!+\!y_3^2
\!+\!2(x_1z_3\!+\!x_2z_3\!+\!x_3z_2\!+\!y_1z_1\\
~+y_2z_2\!+\!y_3z_1)]
+J [(x_1\!+\!y_1)^2\!+\!(x_2\!+\!y_2)^2\!+\!z_3^2\!+\!2(z_1^2\!+\!z_2^2\\
~-x_1y_3\!-\!x_2x_3\!+\!x_3y_1\!+\!y_2y_3\!-\!z_1z_3\!-\!z_2z_3\!-\!x_3y_3)] \Big\} \\
~-\sfrac{1}{6}\sqrt{2}S\mu_B H g_{cc}(x_1\!-\!x_2\!+\!x_3\!+\!y_1\!-\!y_2\!-\!y_3) \,,
\end{array}
\see
or, in terms of the structure factor components, 
\sbe\label{eq:energy_c2}
\renewcommand{\arraystretch}{1.2}
\begin{array}{l}
E_{\bf c}/\mc{N}\!=\!
\eta_{aA} M_a(A)^2 \!+\!\eta_{bC} M_b(C)^2 \!+\!\eta_{cF} M_c(F)^2 
\!+\! \eta_{aG} M_a(G)^2 \\
~+\eta_{bF} M_b(F)^2 \!+\! \eta_{cC} M_c(C)^2
\!+\! \eta'_{aG} M'_a(G)^2 \!+\! \eta'_{bF} M'_b(F)^2 \\
~+ \eta'_{cF} M'_c(F)^2
\!-\! \sqrt{2} \Gamma \big[
M_a(A) M_b(C) \!-\!i M_a(G) M_b(F) \\ 
~+ \sqrt{3} M_b(F) M_c(C) \!+\! \sqrt{3} M_b(C) M_c(F) \!+\! M'_a(G) M'_b(F)
\big] \\
~- \sqrt{3} K \big[ i M_a(G) M_c(C) \!+\! M_a(A) M_c(F) \big] \!-\! g_{cc}\mu_B H M'_c(F)\,.
 \end{array}
\see
where
\sbea
\renewcommand{\arraystretch}{1.2}
\begin{array}{c}
\eta_{aG}=\Gamma+K/2,~~
\eta_{bF}=-(2J+K),~~
\eta_{cC}=-\Gamma+K/2,\\
\eta'_{cF}=\sfrac{1}{2}(\Gamma + 3 J + K)
\end{array}
\seea

\vspace*{-0.3cm}
\subsubsection{High-field phase for {\Hc}}
\vspace*{-0.3cm}
For $H\!\geq\!H_{\bf c}^\ast$, the Cartesian components satisfy the relations
\sbe
\begin{array}{c}
x_1\!=\!-y_2\!=\!-y_3\,,~~
y_1\!=\!-x_2\!=\!x_3\,,~~
z_1\!=\!z_2\!=\!-z_3\,,
\end{array}
\see
and we are left with two spin sublattices [see Table~\ref{tab:ansatze}],  
\sbe
\renewcommand{\arraystretch}{1.2}
\begin{array}{l}
{\bf A}={\bf B}={\bf C}\equiv {\bf F} =S [x_1,y_1,z_1]\,,\\
{\bf A}'={\bf B}'={\bf C}'\equiv {\bf F}' = S [y_1,x_1,z_1]\,,
\end{array}
\see
and one spin length constraint, $x_1^2\!+\!y_1^2+\!z_1^2\!=\!1$. 
Equivalently, all modulated static structure factor components vanish identically, and we are left with the three uniform components $M'_a(G)$, $M'_b(F)$ and $M'_c(F)$, 
\sbe
\renewcommand{\arraystretch}{1.2}
\begin{array}{c}
M_a'(G) = -\frac{S}{\sqrt{2}} (x_1-y_1)\,,~~
M_b'(F) = -S z_1\,,\\[1ex]
M_c'(F) = -\frac{S}{\sqrt{2}} (x_1+y_1)\,,
\end{array}
\see
subject to the constraint $M_b'(F)^2\!+\!M_c'(F)^2\!+\!M_a'(G)^2\!=\!S^2$.  
The total energy Eq.~(\ref{eq:energy_c2}) becomes 
\sbe\label{eq:energy_c3}
\renewcommand{\arraystretch}{1.2}
\begin{array}{l}
E_{\bf c}/\mc{N} \!=\! \eta'_{bF} M_b'(F)^2\!+\! \eta'_{cF}  M_c'(F)^2\!+\! \eta'_{aG}  M_a'(G)^2 \\
~~-\sqrt{2}\Gamma~M_a'(G)~M_b'(F) -g_{cc}\mu_B H ~M_c'(F)\,.
\end{array}
\see

Here the minimization of the energy for $H\!\ge\!H_{\bf c}^\ast$ gives the following relations
\sbe\label{eq:MpaGvsH}
\renewcommand{\arraystretch}{1.3}
\begin{array}{c}
M'_a(G)\!=\!-S \frac{\sqrt{1\!-\!(H/H_{\bf c}^{\ast\ast})^2}}{\sqrt{1\!+\!t^2}},~
M'_b(F)\!=\! -t~M'_a(G),\\
M'_c(F)\!=\!-S H/H_{\bf c}^{\ast\ast}
\end{array}
\see
where $t\!=\!\frac{2J S-g_{cc} H_{\bf c}^{\ast\ast}}{\sqrt{2}\Gamma S}$ and $H_{\bf c}^{\ast\ast}$ is given by Eq.~(\ref{eq:Hcstarstar}).

Note that the cross-coupling term $-\sqrt{2}\Gamma M_a'(G) M_b'(F)$ in Eq.~(\ref{eq:energy_c3}) favors opposite signs for $M_a'(G)$ and $M_b'(F)$, since $\Gamma\!<\!0$. However, unlike the cases {\Ha} and {\Hb}, here none of the signs of $M_a'(G)$ and $M_b'(F)$ are fixed by the field, meaning that the system can spontaneously choose either $M_a'(G)\!>\!0$ and $M_b'(F)\!<\!0$ or $M_a'(G)\!<\!0$ and $M_b'(F)\!>\!0$. The associated broken symmetries are $\Theta C_{2{\bf b}}$ and $C_{2{\bf c}}$, see Table~\ref{tab:symmetries}.

\vspace*{-0.3cm}
\section{Proof of Eqs.~(\ref{eq:FeynmanHellmann})}\label{app:FeynmanHellmann}
\vspace*{-0.3cm}
Here we show a mathematical proof of Eqs.~(\ref{eq:FeynmanHellmann}). The proof is based on a classical version of the so-called Feynman-Hellmann theorem known in Quantum Mechanics. 
We begin by writing the total classical energy of the system as a function  of the spherical coordinates $\{\theta_i,\phi_i\}$ of the spins ($i=1$-$\mc{N}$, the total number of spins) and  the free parameters of the model, namely $J$, $K$, $\Gamma$ and $H$:
\sbe
E_{\text{class}} =  f(\{\theta_i,\phi_i\}; J, K, \Gamma, H).
\see
Let us denote the classical ground state configuration for a given set of $J$, $K$, $\Gamma$ and $H$ by $\{\theta_i^*,\phi_i^*\}$, where
\sbe
\theta_i^*=\theta_i^*(J, K, \Gamma, H)\,,~~~
\phi_i^*=\phi_i^*(J, K, \Gamma, H)\,.
\see
These angles are found by minimizing the total energy
\sbe\label{eq:minthetaphi}
\frac{\partial f}{\partial \theta_i}\Big|_{\theta_i=\theta_i^*,\phi_i=\phi_i^*}=0,~~~
\frac{\partial f}{\partial \phi_i}\Big|_{\theta_i=\theta_i^*,\phi_i=\phi_i^*}=0
\see
 Then the minimum of the classical energy, or  the classical ground state energy,  is given by
\sbea
E_{\text{class,min}} &=& f(\{\theta_i^*,\phi_i^*\}; J, K, \Gamma, H) \equiv E(J, K, \Gamma, H) \nonumber\\
&=& E_J+ E_K+ E_\Gamma+ E_{\text{Z}}\,,
\seea
where the terms in the second line are the individual contributions to the energy from the $J$, $K$ and $\Gamma$ interactions, and the Zeeman field, respectively. 
We can now formulate the classical version of the Feynman-Hellmann theorem by taking the derivative of the ground state energy with respect to the parameter $J$, as an example. We have
\sbe
\frac{\partial E}{\partial J} = \left(\sum_i \left( \frac{\partial f}{\partial \theta_i} \frac{\partial \theta_i}{\partial J} + \frac{\partial f}{\partial \theta_i} \frac{\partial \theta_i}{\partial J} \right)
+ \frac{\partial f}{\partial J}\right)_{\theta_i=\theta_i^*,\phi_i=\phi_i^*}\,,
\see
and using Eqs.~(\ref{eq:minthetaphi}) we get
\sbe
\partial E/\partial J= \left(\partial f/\partial J\right)_{\theta_i=\theta_i^*,\phi_i=\phi_i^*} = E_J/J\,,
\see
where in the last step we used the fact that $f(\{\theta_i^*,\phi_i^*\}; J, K, \Gamma, H)$ depends linearly on $J$.
Similarly, for the other free parameters we get
\sbe
\frac{\partial E}{\partial K} = \frac{E_K}{K}\,,~~~
\frac{\partial E}{\partial \Gamma} = \frac{E_\Gamma}{\Gamma}\,,~~~
\frac{\partial E}{\partial H} = \frac{E_{\text{Z}}}{H} =  - \mc{N} m_\parallel\,,~~~
\see 
where $m_\parallel$ is the magnetization per site along the field.

To arrive at Eqs.~(\ref{eq:FeynmanHellmann}) we need to look at the second derivatives of $E$. For example, 
\sbea
\frac{\partial^2 E}{\partial J \partial H} &=& \frac{\partial}{\partial J}\left(  \frac{\partial E}{\partial H} \right) =  -\mc{N}  \frac{\partial m_\parallel}{\partial J}\,, \\
\frac{\partial^2 E}{\partial H \partial J} &=& \frac{\partial}{\partial H}\left(  \frac{\partial E}{\partial J} \right) = \frac{1}{J} \frac{\partial E_J}{\partial H} \,.
\seea
The equality $\frac{\partial^2 E}{\partial J \partial H}\!=\!\frac{\partial^2 E}{\partial H \partial J}$ then gives
\sbe
-\mc{N}  \frac{\partial m_\parallel}{\partial J} =  \frac{1}{J} \frac{\partial E_J}{\partial H}\,,
\see
and similarly for the remaining equations of (\ref{eq:FeynmanHellmann}).

\vspace*{-0.3cm}
\section{Spin wave analysis and reduction of sublattice magnetizations due to quantum fluctuations }\label{app:SW} 
\vspace*{-0.3cm}
In this Appendix we provide the details for the semiclassical expansion around the classical ans\"atze of Table~\ref{tab:ansatze} and the calculation of the total magnetization for all field directions. We shall only discuss the case of the six-sublattice states for $H\!<\!H^\ast$. The analysis of the high-field two-sublattice states follows along the same lines.

\begin{figure}[!t]
\includegraphics[width=0.95\columnwidth,trim= 0 0 0 0,clip]{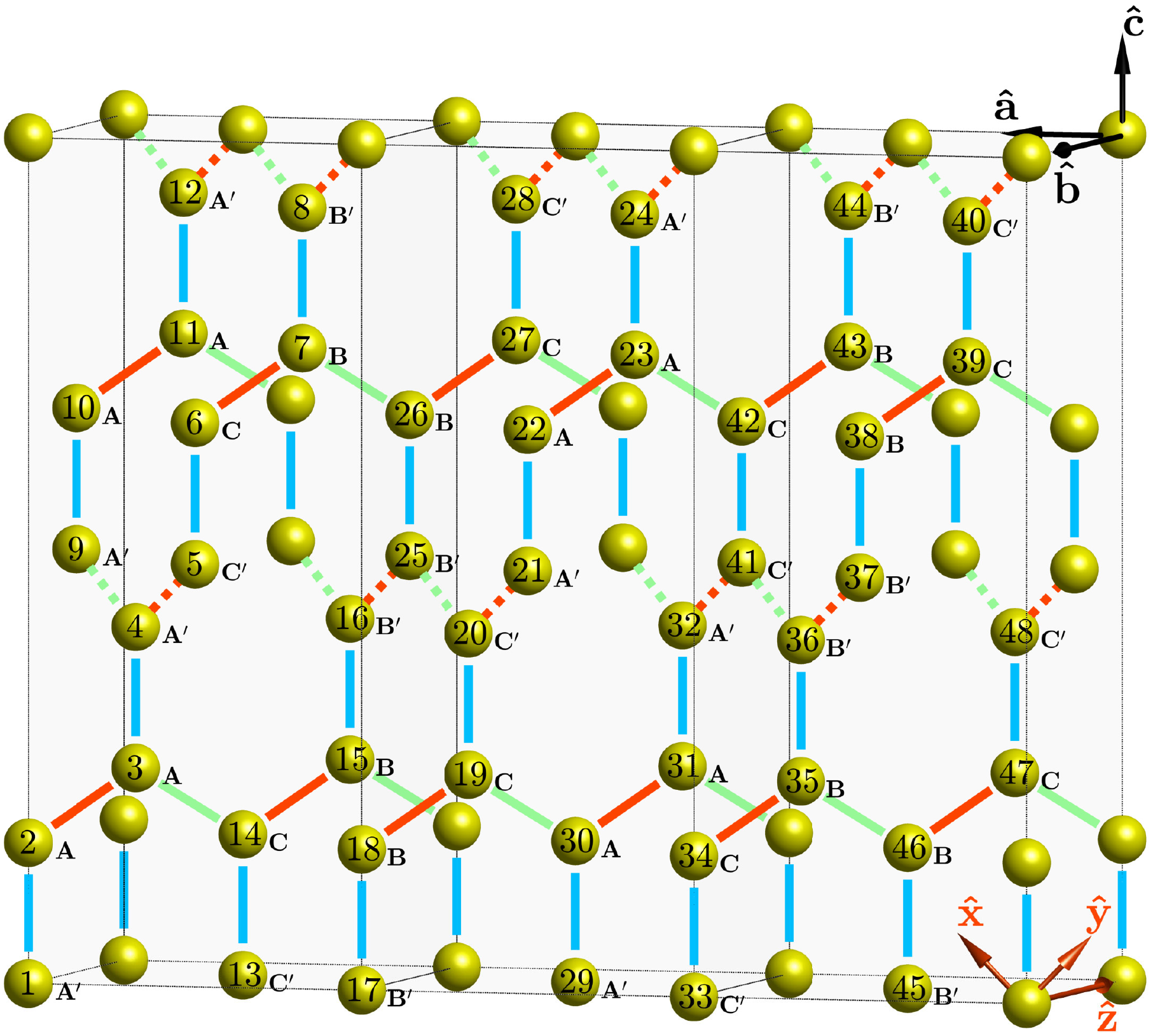}
\caption{A sketch of a hyperhoneycomb lattice with three orthorhombic unit cells, which is also a magnetic unit cell in a $K$-state, where the six sublattices along the $xy$ and $x'y'$ chains are labeled as $\bf A$, $\bf B$, $\bf C$, and $\bf A'$, $\bf B'$, $\bf C'$, respectively.}\label{fig:MUC}
\end{figure}

\vspace*{-0.3cm}
\subsection{Quadratic spin-wave Hamiltonian}
\vspace*{-0.3cm}
We first relabel the spin sites as $i\!\to\!({\bf R},\mu)$, where now ${\bf R}\!=\!(3 n_1, n_2, n_3)$ denotes the position of the magnetic unit cell in the orthorhombic frame ($n_1$, $n_2$, and $n_3$ are integers), and $\mu\!=\!1$-$\mc{N}_{\text{m}}$ is the sublattice index inside the magnetic unit cell, with $\mc{N}_{\text{m}}\!=\!48$. The magnetic cell and the corresponding labeling convention is shown in Fig.~\ref{fig:MUC}. 
To proceed we rewrite the spins as ${\bf S}_i \!\to \!{\bf S}_{{\bf R}, \mu}$ and their physical positions as $\vec{r}_i =\vec{R}+\bs{\rho}_\mu$, where $\bs{\rho}_\mu$ is the sublattice vector associated with $\mu$-th sublattice. The Hamiltonian (\ref{eq:Hamiltonian}) is then written as 
\sbe
\mc{H}=\frac{1}{2}\sum_{{\bf R},\mu,\nu}{\bf S}_{{\bf R},\mu}^{\text{T}} \cdot \bs{\mc{J}}_{\mu\nu} \cdot {\bf S}_{{\bf R}+{\bf t}_{\mu\nu}, \nu}
-\mu_B H\sum_{{\bf R},\mu}\bs{g}_{{\bf R},\mu}^{\text{T}} \cdot {\bf S}_{{\bf R},\mu}\,,
\see
where ${\bf S}_{{\bf R},\mu}^{\text{T}}=(S_{{\bf R},\mu}^x,S_{{\bf R},\mu}^y,S_{{\bf R},\mu}^z)$, ${\bf t}_{\mu\nu}$ is a primitive translation of the superlattice such that the spins at sites $i=({\bf R},\mu)$ and $j=({\bf R}+{\bf t}_{\mu\nu},\nu)$ interact with each other via $\bs{\mc{J}}_{\mu\nu}$, and 
\sbe
\bs{\mc{J}}_{\mu\nu}=\left\{
\begin{array}{rcl}
\bs{\mc{J}}_{t}, & & \text{if}~~\bs{\rho}_{\mu}-\left({\bf t}_{\mu\nu}+\bs{\rho}_\nu\right) = \pm\bs{\delta}_t, \\ 
0, & & {\text{otherwise}}
\end{array}
\right.\,,
\see
where $\bs{\delta}_t$ connects NN spin sites sharing a bond of type $t \in \{x, y, z, x', y' \}$ (see Sec.~\ref{sec:LatticeSymmetry} and Fig.~\ref{fig:lattice}), and 
\sbea
&&\bs{\mc{J}}_{x}=\begin{pmatrix}
{J\!+\!K} & ~~ 0~~  & ~~ 0~~\\
~~ 0~~ & J & \Gamma\\
~~ 0~~ & \Gamma & J\\
\end{pmatrix},\,
\bs{\mc{J}}_{y}=\begin{pmatrix}
J & ~~ 0~~ & -\Gamma\\
~~ 0~~ & {J\!+\!K} & ~~ 0~~\\
-\Gamma & ~~ 0~~ & J\\
\end{pmatrix},\nn\\
&&\bs{\mc{J}}_{z}=\begin{pmatrix}
J & \Gamma & ~~ 0~~\\
\Gamma & J & ~~ 0~~\\
~~ 0~~ & ~~ 0~~ & {J\!+\!K}
\end{pmatrix},\, 
\bs{\mc{J}}_{x'}=\begin{pmatrix}
{J\!+\!K} & ~~ 0~~ & ~~ 0~~\\
~~ 0~~ & J & -\Gamma\\
~~ 0~~ & -\Gamma & J
\end{pmatrix},\nn\\
&&\bs{\mc{J}}_{y'}=\begin{pmatrix}
J & ~~ 0~~ & \Gamma\\
~~ 0~~ & {J\!+\!K} & ~~ 0~~\\
\Gamma & ~~ 0~~ & J
\end{pmatrix},\, 
\bs{g}_{{\bf R},\mu}=\begin{pmatrix}
\sfrac{1}{\sqrt{2}} \zeta_{{\bf R},\mu} g_{ab} \\
-\sfrac{1}{\sqrt{2}} \zeta_{{\bf R},\mu} g_{ab} \\
-g_{bb}
\end{pmatrix}.
\seea
Here, in order to describe the staggered  nature of the $g$-factor, we  denote $\zeta_{{\bf R},\mu}=1$ for $\vec{R}+\bs{\rho}_{\mu}\in xy$ chain and $\zeta_{{\bf R},\mu}=-1$ for $\vec{R}+\bs{\rho}_{\mu}\in x'y'$ chain. 

Next, for each site $i=({\bf R},\mu)$, we introduce the local reference frame $\{\widetilde{\vec{x}}_i,\widetilde{\vec{y}}_i,\widetilde{\vec{z}}_i\}$ such that $\widetilde{\vec{z}}_i$ coincides with the direction of spin ${\bf S}_i$ in the classical ground state. The spin is then rotated into this local frame of reference by $\widetilde{\vec{S}}_{{\bf R},\mu}=\vec{U}_\mu \cdot \vec{S}_{{\bf R},\mu}$, where the unitary rotation matrix $\vec{U}_\mu$ can be constructed using the polar and azimuthal angles $(\theta_\mu,\phi_\mu)$ associated with the direction of the spin in the classical ground state, 
\sbe
\vec{U}_\mu=\begin{pmatrix}
\cos\theta_\mu\cos\phi_\mu\quad & \cos\theta_\mu\sin\phi_\mu\quad & -\sin\theta_\mu\\
-\sin\phi_\mu\quad & \cos\phi_\mu\quad & 0\\
\sin\theta_\mu\cos\phi_\mu\quad & \sin\theta_\mu\sin\phi_\mu\quad & \cos\theta_\mu
\end{pmatrix}.
\see
Subsequently, we express the local spins in terms of the Holstein-Primakoff  bosons  $a_{{\bf R},\mu}^{\dagger}$ and $a_{{\bf R},\mu}$  and expand the Hamiltonian in powers of $1/\sqrt{S}$ about the classical limit. Collecting the terms that are quadratic in the bosonic operators and going into momentum space, with $a_{{\bf q},\mu}\!=\!\frac{1}{\sqrt{\mc{N}_{\text{m}}}}\sum_{\bf R} e^{i {\bf q}\cdot{\bf R}} a_{{\bf R},\mu}$ (with ${\bf q}$ belonging to the first magnetic Brillouin zone) gives 
\sbe
\mc{H}_2=E_{cl}/S+\sum_{\vec{q}} {\bf x}_\vec{q}^\dagger \cdot {\bf H}_{\vec{q}}\cdot {\bf x}_\vec{q}\; ,
\see
where $E_{cl}$ is the classical energy, 
\sbe
{\bf x}_\vec{q}=\left( a_{\vec{q},1}\, ,...\, ,a_{\vec{q},\mc{N}_{\text{m}}}\, ,a_{-\vec{q},1}^\dagger\, ,...\, ,a_{-\vec{q},\mc{N}_{\text{m}}}^\dagger \right)^{\text{T}}\,,
\see 
where ${\bf H}_{\vec{q}}$ is a $(2\mc{N}_{\text{m}})\times(2\mc{N}_{\text{m}})$ matrix. The diagonalization of ${\bf H}_\vec{q}$ involves introducing a set of Bogoliubov quasiparticle operators~\cite{Bogoliubov1947,Blaizot} 
\sbe
{\bf y}_\vec{q}\!=\!\left( b_{\vec{q},1}\, ,...\, ,b_{\mc{N}_{\vec{q},\text{m}}}\, ,b_{-\vec{q},1}^\dagger\, ,...\, ,b_{-\vec{q},\mc{N}_{\text{m}}}^\dagger \right)^{\text{T}}\,, 
\see 
obtained from ${\bf x}_\vec{q} $ by a unitary canonical transformation ${\bf x}_\vec{q}=\vec{T_q}\cdot {\bf y}_\vec{q}$, where $\vec{T_q}$  satisfies the bosonic commutation relations $\vec{T_q}^\dagger\cdot\bs{\eta}\cdot\vec{T_q}=\bs{\eta}$, with $\bs{\eta}=\text{diag}(\vec{I},-\vec{I})$ and $\vec{I}$ is a $\mc{N}_{\text{m}}\times \mc{N}_{\text{m}}$ unitary matrix.
The matrix ${\bf T}_{\bf q}$ can be found by solving the eigenvalue equation~\cite{Blaizot} $(\bs{\eta}\cdot{\bf H}_{\bf q})\cdot {\bf T}_{\bf q}={\bf T}_{\bf q}\cdot (\bs{\eta}\cdot\bs{\Omega}_{\bf q})$, where  
\sbe\label{spectrum}
\bs{\Omega}_\vec{q}=\vec{T_q}^\dagger \cdot {\bf H}_{\vec{q}} \cdot \vec{T_q}=\text{diag}(\bs{\omega}_{\vec{q}},-\bs{\omega}_{\vec{q}}) \; ,
\see
and $\bs{\omega}_{\vec{q}}=\text{diag}\left(\omega_{\vec{q},1}, \omega_{\vec{q},2}, \ldots,\omega_{\vec{q},\mc{N}_{\text{m}}}\right)$ contains the frequencies of the elementary magnon excitations.

\vspace*{-0.3cm}
\subsection{Total magnetization and torque at zero temperature}\label{app:mandtau}
\vspace*{-0.3cm}
To find the total magnetization of the system we must first compute the expectation values of the spins in the local frame. 
To leading order in the semiclassical expansion we have
\sbe
\langle \widetilde{S}_{{\bf R},\mu}^x\rangle \simeq 0\,,~~~
\langle \widetilde{S}_{{\bf R},\mu}^y\rangle \simeq 0\,,
\see
while symmetry dictates that 
\sbe
\langle \widetilde{S}_{{\bf R},\mu}^z\rangle = S-\langle a^\dagger_{{\bf R},\mu} a_{{\bf R},\mu}\rangle \equiv S- \zeta_\mu = \text{independent of}~{\bf R}\,.
\see
This property allows to rewrite the spin length reduction $\zeta_\mu$ as
\sbe
\zeta_\mu = \frac{\mc{N}_{\text{m}}}{\mc{N}} \sum_{\bf R} \langle a^\dagger_{{\bf R},\mu} a_{{\bf R},\mu}\rangle = \frac{\mc{N}_{\text{m}}}{\mc{N}} \sum_{\bf q} \langle a^\dagger_{{\bf q},\mu} a_{{\bf q},\mu}\rangle\,,
\see
where ${\bf q}$ belongs to the first magnetic Brillouin zone and the total number of magnetic unit cells is given by $\mc{N}/\mc{N}_{\text{m}}$. 
Using the $T\!=\!0$ limit of the standard relations 
\sbe
\langle b_{\vec{q},i}^\dagger b_{\vec{q},j}\rangle=\delta_{ij} ~n(\omega_{\vec{q},i})\,,~~~
\langle b_{\vec{q},i} b_{\vec{q},j}^\dagger\rangle=\delta_{ij} ~[1+n(\omega_{\vec{q},i})] \,,
\see
where $n(\omega)\!=\!\big(e^{\hbar\omega/(k_BT)}\!-\!1\big)^{-1}$ is the Bose-Einstein distribution function, we arrive at the zero-temperature expression for $\zeta_\mu$:
\sbe
\zeta_{\mu} = \frac{\mc{N}_{\text{m}}}{\mc{N}} \sum_{{\bf q}} \sum_{j=\mc{N}_{\text{m}}+1}^{2\mc{N}_{\text{m}}} 
\left({\bf T}_{\bf q}\right)_{j\mu}^\ast \!\left({\bf T}_{\bf q}\right)_{\mu j}\,.
\see
Next we use the relation $\vec{S}_{{\bf R},\mu}=\vec{U}_\mu^{-1} \cdot \widetilde{\vec{S}}_{{\bf R},\mu}$, where 
\sbe
\vec{U}_\mu^{-1}=\begin{pmatrix}
\cos\phi_\mu\cos\theta_\mu\quad & -\sin\phi_\mu\quad & \cos\phi_\mu\sin\theta_\mu\\
\sin\phi_\mu\cos\theta_\mu\quad & \cos\phi_\mu\quad & \sin\phi_\mu\sin\theta_\mu\\
-\sin\theta_\mu\quad & 0\quad & \cos\theta_\mu
\end{pmatrix}\,,
\see
to arrive at
\sbe
\begin{array}{l}
\langle S_{{\bf R}, \mu}^a \rangle \equiv \langle S_{\mu}^a \rangle = \frac{\sin\theta_\mu}{\sqrt{2}} \left( \cos\phi_\mu-\sin\phi_\mu\right) \left( S-\zeta_\mu\right)\,, \\[1.2ex]
\langle S_{{\bf R}, \mu}^b \rangle \equiv \langle S_{\mu}^b \rangle= -\cos\theta_\mu \left( S-\zeta_\mu\right)\,, \\[1.2ex]
\langle S_{{\bf R}, \mu}^c \rangle \equiv \langle S_{\mu}^c \rangle= \frac{\sin\theta_\mu}{\sqrt{2}} \left( \cos\phi_\mu+\sin\phi_\mu\right) \left( S-\zeta_\mu\right)\,.
\end{array}
\see
which are all independent of ${\bf R}$. 
Having computed the expectation values $\langle {\bf S}_\mu\rangle$ we can then compute the magnetization per site ${\bf m}$ using Eq.~(\ref{eq:magn}) of the main text, while the torque per site is given by $\bs{\tau} = {\bf m} \times {\bf H}$.

\vspace*{-0.3cm}
\section{Monte Carlo simulation}\label{app:MC}
\vspace*{-0.3cm}
 In this Appendix we present some details of  the classical Monte Carlo  (MC) simulations which we   employed for calculating the finite temperature phase diagram of the model  (\ref{eq:Hamiltonian}) similar to Refs.~[\onlinecite{Price2012,Price2013,Chern2017}]. In our simulations, we treat the spins as three-dimensional vectors, $\vec{S}=(S_x,S_y,S_z)$, of unit magnitude with $S^2_x+S_y^2+S_z^2=1$. To ensure a uniform sampling, we first generate two random numbers $r_1$ and $r_2$ which are both uniformly distributed on $(0,1)$~\cite{Marsaglia1972}. Then we have $S_z=2r_1-1$, $S_x=\sqrt{1-S_z^2}\cos(2\pi r_2)$, and $S_y=\sqrt{1-S_z^2}\sin(2\pi r_2)$. The simulations were performed on different systems with a total number of sites equal to ${\mc N}\!\in\!\{48,96,144,192,240,288\}$.  At each temperature, more than 10$^6$ MC sweeps were performed. Of these, 10$^5$ MC sweeps were used to calculate the averages of physical quantities. 

\begin{figure}[!t]
{\includegraphics[width=0.9\columnwidth]{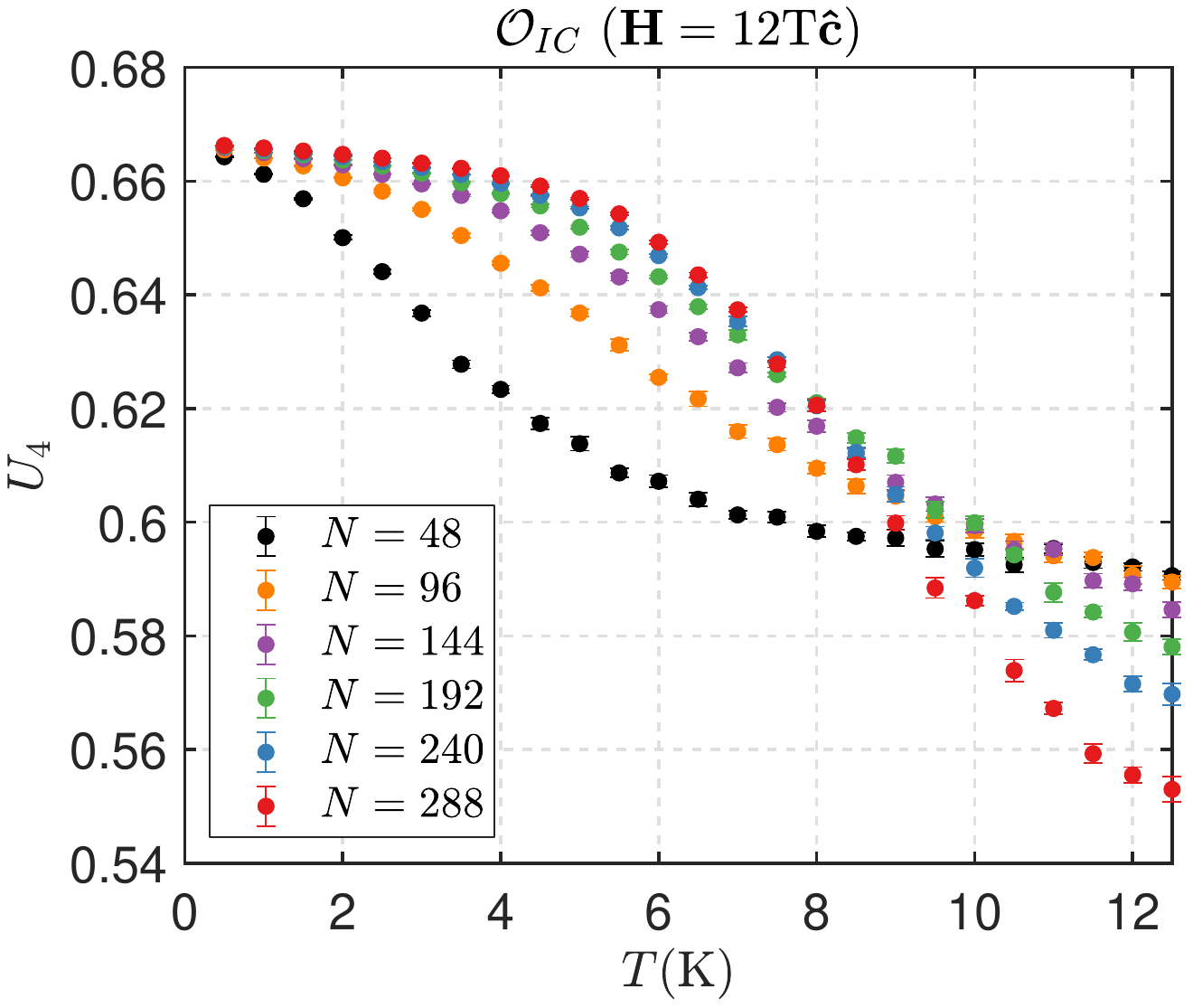}}
\caption{ Binder's cumulants  as functions of $T$  computed for ${\mc N}\in\{48,96,144,192,240,288\}$  for the  $\bf c$ magnetic field of the magnitude  $H_{\bf c}=12$T.  The  errors are  calculated from a Jackknife binning analysis.}  \label{fig:binder}
\end{figure}

To reduce the autocorrelation time, we have used the standard Metropolis algorithm combined with the over-relaxation algorithm~\cite{Metropolis1953,Creutz1987}. Namely, one Metropolis sweep was performed after completing ten over-relaxation sweeps where each sweep contains ${\mc N}$ updates. The over-relaxation process with single spin updates is given by~\cite{Creutz1987}, 
\sbe
\vec{S}'_{{\bf R},\mu} = -\vec{S}_{{\bf R},\mu} + 2 \frac{\vec{S}_{{\bf R},\mu}\cdot\vec{h}_{{\bf R},\mu}}{|\vec{h}_{{\bf R},\mu}|^2}\vec{h}_{{\bf R},\mu}\,,
\see
where ${\bf h}_{{\bf R},\mu}$ is the local effective field at  site $i=({\bf R},\mu)$. Compared with other MC updates, over-relaxation usually costs less computing time and has less autocorrelations. However, because the over-relaxation update is a micro-canonical process, we adopt the standard Metropolis algorithm to ensure ergodicity of the simulation. In each Metropolis update, one spin $\vec{S}_{{\bf R},\mu}$ is randomly chosen and altered to a new direction confined within a cone defined by $d\theta \in [0,\pi]$. We first rotate the coordinate such that the $z$-axis coincides with $\vec{S}_{{\bf R},\mu}$, i.e. the center of the cone. Similar to the initialization process,  here again we generate two random numbers $p_1$ and $p_2$ in the interval $(0,1)$ and take 
\sbea
&&S_z=\cos(d\theta)+[1-\cos(d\theta)]p_1 \; , \nn\\
&&S_x=\sqrt{1-S_z^2}\cos(2\pi p_2) \; , \nn\\
&&S_y=\sqrt{1-S_z^2}\sin(2\pi p_2) \; ,
\seea
at which point we generate a random, uniformly distributed unit vector $[S_x,S_y,S_z]$ within the cone. Afterwards, we rotate the coordinate back to the original coordinate and compute the change in energy $\Delta E$ which is related to the probability of acceptance.
In the equilibration process, which is the transient time for the system to reach equilibrium, we gradually adjust the magnitude of $d\theta$ such that the acceptance ratio keeps staying within $[0.4,0.6]$. In the measurement process, we take one measurement of the observables after every ten Metropolis sweeps.

Next, in order to obtain the critical temperatures, we  have used the Binder cumulants method.	The  fourth-order Binder  cumulant, $U_4=1-\frac{\langle{\mc O}^4 \rangle}{3\langle{\mc O}^2 \rangle^2}$, where  $ {\mc O}$ denote some long-range order parameter, has a scaling dimension of zero; thus the crossing point of the cumulants for different lattice sizes provides a reliable estimate for the value of the critical temperature  $T_c$ at which the long range order is destroyed. In Fig.~\ref{fig:binder},  we plot the Binder's cumulants for ${\mc N}\in\{48,96,144,192,240,288\}$  for the case when magnetic field  with the magnitude  $H_{\bf c}=12$T is applied along the ${\bf c}$ crystallographic axis. The corresponding statistical errors are calculated using a Jackknife binning analysis with ten bins~\cite{Efron1982}.

\bibliographystyle{apsrev4-1}
\bibliography{reference}

\end{document}